\documentclass[10pt]{article}
\usepackage[english]{babel}

\usepackage{amsmath}
\usepackage{amssymb}
\usepackage{amsfonts}
\usepackage{esint}
\usepackage{graphicx}
\usepackage{subfigure}
\usepackage{refcheck}
\usepackage{stmaryrd}
\usepackage{color}
\usepackage{epstopdf}
\usepackage{cite}
\usepackage{hyperref}
\usepackage{tikz}
\usepackage{stmaryrd}

\graphicspath{{figures/}} 

\begin{document}

\title{Hidden ring crack in a rotating cylindrical shell under torsion}

\author{Z. Zhuravlova$^{(1)}$, I. Istenes$^{(2)}$, D. Peck$^{(3)}$, Yu. Protserov$^{(1)}$, N. Vaysfeld$^{(4)}$\\[4mm]
{\it $^{(1)}$ Odessa I.I. Mechnikov National University, Faculty of Mathematics, Physics}\\ {\it and Information Technologies, str. Dvoryanskaya, 2, 65082, Odessa, Ukraine.}
\\[1mm]
{\it $^{(2)}$ ROEZ R{\&}D, Bratislava, Slovakia.}
\\[1mm]
{\it $^{(3)}$ Aberystwyth University, Aberystwyth, United Kingdom.}
\\[1mm]
{\it $^{(4)}$ King's college, Strand building, S2.35, London, United Kingdom.}\\[4mm]
{\footnotesize {\bf E-mails:} z.zhuravlova@onu.edu.ua, igor.istenes@roez.sk, dtp@aber.ac.uk, protserov@onu.edu.ua, natalya.vaysfeld@kcl.ac.uk. }}

\maketitle

\abstract{We consider the impact of a ring crack within a rotating hollow cylinder of fixed height under axisymmetric (torsion) loading. The form of the displacement is obtained from the equation of motion using the Fourier sin transform. The displacement jump over the crack is obtained from the boundary condition on the tangential stress, formulated as a singular integral equation which is solved by the method of orthogonal polynomials. The stress intensity factors on the opposing crack surfaces are calculated. The dependence of the crack extension on the problem geometry is investigated, including the impact of the crack’s location, cylinder’s height, torsion loading and rotation frequency. Possible extensions of the model to cover fatigue cracking are considered. A practical test to detect and locate cracks within a rotating cylinder is outlined.}

\section{Introduction}

Rotational motion is a fundamental aspect of many technological systems, as it enables reliable, efficient conversion of energy and precise control of movement, amongst others. As a result, it forms the basis of many types of machinery, such as engines, turbines, and motors. It is often used when converting energy from one form to another, be it motion, mechanical work, or enabling power generation. Manufacturing processes like drilling, milling and lathe operations rely on controlled rotational motion to shape and modify the materials that shape our modern world.

It is therefore unsurprising that the study of general problems involving rotational motion, and the key features underlying them, remains an important area of modelling and study. The results of studies of the general laws of rotational motion in the theory of shells \cite{ANTMAN2010991}, the behaviour of waves under a rotational load in coupled fields \cite{OTHMAN2008639}, and the nonlinear dynamics of structures \cite{DASTJERDI2020103371}, underlie the modelling of special problems that arise in robotics (simulation of rotational movements of human limbs), energy (wind turbines converting wind energy into electricity using rotational movements), biomechanics (for modelling human joints, muscular mechanisms), and many more.

One of the simplest three-dimensional shapes to consider in such torsion problems is that of a cylinder, which has immediate applications to approximate more complex structures such as pipes, tubes, columns, shafts, and so forth, that are affected by stress and strains. Modern studies of the wave fields of finite cylinders include both the analysis of bodies comprised of different materials (for example, models of orthotropic bi-directional FGMaterials \cite{BATRA2018336}, hyperelastic materials \cite{ENGIN1978387}, inter-layer fracture of carbon nanotubes \cite{VIET2012256}, etc), and their behaviour under various dynamic modes of loading  \cite{GEORGE1978109,LI1995447}. One of the essential factors affecting the stress state of the cylinder is the nature of the applied load. For example, it was shown in \cite{PRONINA2023103936} that for a fixed wall thickness and stress gradient the gain in the ultimate loading capacity depends on the magnitude of the gradient, however it is only weakly dependent on the gradient direction and the pipe radii. Meanwhile, in \cite{WANG2021103432} a weakly nonlinear analysis was conducted for localized necking of a hyperelastic solid cylinder under axial stretching, based on the exact theory of nonlinear elasticity.

In the case where a dynamic load is applied, modelling becomes much more complicated and requires new mathematical approaches. In \cite{SLEPYAN2022103628}, the impact of forced waves in a uniform waveguide with distributed and localized dynamic structures was considered, and the general patterns obtained. Meanwhile, in the context of mixture theory \cite{DOUHOU2021103575} considered the steady diffusion of an ideal fluid through a two-layer thick walled pre-stressed and fibre-reinforced hollow cylinder.

A special case of problems in rotating cylinders is when a small defect is present within the body, as it significantly increases the risk of crack development and the appearance of dangerous stresses. The importance of microstructure modelling for additively manufactured metal post-process simulations was demonstrated in \cite{SUNNY2021103515}. Meanwhile, in \cite{LI1995447} the issue of stress concentration near various types of crack within a cylinder was considered. This was achieved using the torsion problem for a circular cylinder containing an equilateral triangle opening and a line crack, with the solution obtained using the method of singular integral equations and the crack-cutting technique. The case of an external crack on a hollow cylinder under axisymmetric torsion has been considered by numerous authors, such as the case of a circumferential crack in \cite{GAO20082155}, multiple cracks \cite{Akiyama2001137}, and that of a lone crack in \cite{KAZUYOSHI1978707} (see also references therein), although only the latter accounted for cylinder rotation. In that paper, the mixed boundary value problem being reduced to a pair of dual series equations. They found that the magnitude of the stress intensity factor was heavily dependent on the crack location. Similarly, in \cite{ABAKAROV2022103617} they demonstrated that small perturbations away from the case of a symmetric crack configuration (under pure mode I or II) produces mixed-mode conditions at the crack tips. This in turn may lead to an increase in the stress intensity factors and the energy release rates. The case of internal cracks within the cylinder has been studied, for example for a single crack \cite{Duo1994}, multiple cracks \cite{VAYSFELD2017526}, and a numerical approach that can handle multiple internal and external cracks \cite{NOROOZI2018811}, however these approaches did not incorporate potential rotation of the cylinder.

In this paper, we seek to develop a general model for a rotating hollow cylinder of fixed height, containing an internal crack, under axisymmetric torsion. With the above results in mind, we take the problem geometry to allow a thorough investigation of the crack location on the stress intensity factors. The primary aim is to develop ensure the results can be utilized in practical application to the operation of turbines, and other machinery with rotating components. While for the case of turbines it is typically fracture of the blades that is the primary concern (see for example: \cite{CHOWDHURY2023100329} for gas turbines, \cite{Wang2022} for wind turbines), damage to the central shaft can be more difficult to detect - even with routine inspection (for details of proposed inspection methodology, see e.g. \cite{Rau2016,Nurbanasari2014,Xie2020}), and may interfere with indirect damage monitoring methods, such as vibration detection (see e.g. \cite{Liu2017,Techy2022} and references therein). Fractures within the central shaft however allow the possibility of local repair, provided they are detected. The presented model is therefore designed to investigate whether a practical test for detecting and locating cracks within cylindrical shafts (without stopping operation), as well as predicting their rate of extension (fast or fatigue), are possible, and to facilitate any related risk management.

The paper is organised as follows. The problem formulation is given in Sect.~\ref{Sect:ProbForm}, with the cylinder geometry outlined, while the governing equations are stated and normalized. The form of the displacement in the cylinder is obtained in Sect.~\ref{Sect:Displacement} using the Fourier sin transform. The jump of the displacement over the crack follows from expressing the boundary condition on the tangential stress in the form of a singular integral equation, which is then resolved using the method of orthogonal polynomials. This is used to obtain the stress intensity factor on the crack surfaces in Sect.~\ref{Sect:SIFs}. With the solution obtained, in Sect.~\ref{Sect:NumInvest} results are presented for the case of a small steel cylinder. First, the extension of an existing crack is examined in Sect.~\ref{Sect:CrackGrowth}, with an investigation of the influence of the crack and cylinder geometry on quasi-static fracture growth. 
Extensions of the model to examine fatigue cracking are given. Meanwhile, in Sect.~\ref{Sect:Detect} the impact of the crack on the displacement within the cylinder is investigated. Whether a test to determine the presence and location of a crack within a cylinder can be created from the presented formulation is investigated. Finally, concluding remarks are given in Sect.~\ref{Sect:Concl}.

\section{Problem Formulation}\label{Sect:ProbForm}

\subsection{Governing equations}

\begin{figure}[h]
 \centering
 \includegraphics[width=0.5\textwidth]{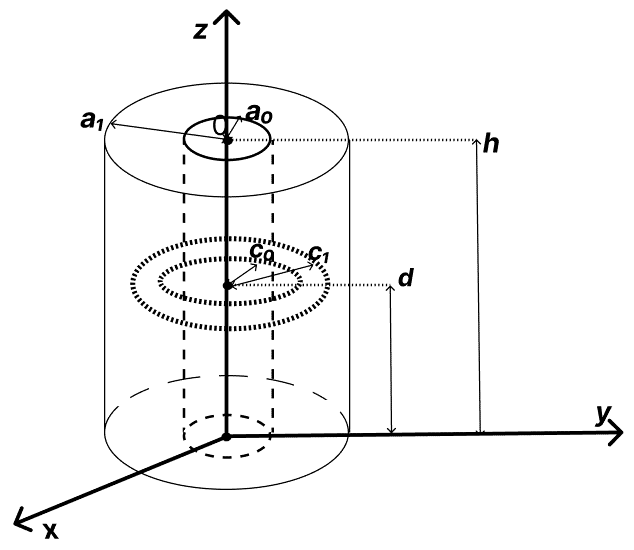}
 \caption{Geometry and coordinate system for a cylinder with a circular crack.}
 \label{Fig:Geometry}
\end{figure}

We consider a hollow elastic cylinder containing a ring crack in cylindrical coordinates (see Fig.~\ref{Fig:Geometry}). The cylinders inner radius is a distance $a_0$ from the origin, its outer radius a distance $a_1$, and has height $h$. The ring crack is located at height $d$, with inner radius $c_0$ and outer radius $c_1$. The problem domain is therefore, in cylindrical coordinates $(R,\phi ,Z)$, given by: $a_0 < R < a_1$, $-\pi < \phi < \pi$, $0<Z<h$. The cylinder is rotating with frequency $\tilde{\omega}$, while the medium has wavespeed $c$. Additional axisymmetric loading (torsion) is applied on the inner, $P_0 (z)$, and outer, $P_1(z)$, surfaces of the cylinder.

The equation of motion takes the form
\begin{equation} \label{Eqn_Motion_1}
\frac{1}{R} \frac{\partial}{\partial R}\left( R \frac{\partial u}{\partial R} \right) - \frac{1}{R^2} u + \frac{\partial^2 u}{\partial Z^2} = -\frac{\tilde{\omega}^2}{c^2} u , \quad a_0 < R < a_1, \quad 0 < Z < h ,
\end{equation}
where here $u = u_\phi (R,Z)$ is the displacement, which will have a discontinuity over the crack surfaces.

The cylinder is assumed to be fixed at the bottom edge
\begin{equation}\label{fixed_edge_1}
u(R,0) = 0, \quad a_0 < R < a_1.
\end{equation}
The upper edge of the cylinder is free from loading
\begin{equation}\label{fixed_stress_1}
\tau_{z\phi} (R, h) = 0 , \quad a_0 < R < a_1,
\end{equation}
while the cylindrical boundaries are under the tangential loading
\begin{equation}\label{loading_1}
\tau_{r\phi} (a_i , Z) = P_i (Z), \quad 0 < Z < h, \quad i=0,1,
\end{equation}
where $\tau_{z\phi} (R,Z)$, $\tau_{r\phi}(R,Z)$ are the tangential stresses, while $P_i(Z)$ is a prescribed (known) function.

Inside the cylinder, the crack results in a displacement jump (denoted by double brackets $\llbracket . \rrbracket$) 
\begin{equation} \label{displacement_jump_1}
 \llbracket u(R,d) \rrbracket = \tilde{\chi} (R), \quad c_0 < R < c_1,
\end{equation}
where the jump $\llbracket u(R,d) \rrbracket = u(R, d - 0 ) - u(R, d + 0)$, while $\tilde{\chi} (R)\neq 0$ is an unknown jump function to be computed as part of the solution. The tangential stress over the crack is such that
\begin{equation}\label{tangential_jump_1}
\llbracket \tau_{z\phi} (R,d) \rrbracket  = 0 , \quad c_0 < R < c_1.
\end{equation}

\subsection{Normalization}

We introduce the following normalization
\begin{equation}\label{Normalization}
r = \frac{R}{a_1 - a_0} , \quad z = \frac{Z}{h} , \quad w(r,z) = \frac{u(R, Z)}{a_1 - a_0} , 
\end{equation}
$$
p_i (z) = \frac{ P_i (Z)}{(a_1 - a_0 ) G F} , \quad \omega = \frac{\tilde{\omega}}{\Omega}, \quad \chi(r) = \frac{\tilde{\chi}(R)}{a_1 - a_0},
$$
where $G$ is the shear modulus, $F$ is the maximal applied load, and $\Omega$ is the maximal frequency.

Under this normalization, the problem \eqref{Eqn_Motion_1} -- \eqref{tangential_jump_1} can be expressed in the form
\begin{equation} \label{Problem_Formulation_1}
\left\{ 
\begin{array}{l} 
\displaystyle\frac{1}{r} \displaystyle\frac{\partial}{\partial r}\left(r \displaystyle\frac{\partial w}{\partial r} \right) - \displaystyle\frac{1}{r^2} w + \gamma^2 \displaystyle\frac{\partial^2 w}{\partial z^2} = -\Psi^2 w , \quad \rho_0 < r < \rho_1 , \quad 0 < z < 1, \\[2mm]
w(r,0) = 0 , \quad \rho_0 < r < \rho_1 , \\[2mm]
\displaystyle\frac{\partial w}{\partial z}(r,1) = 0 , \quad \rho_0 < r < \rho_1,\\[2mm]
\rho_i \displaystyle\frac{\partial w}{\partial r}(\rho_i, z) - w (\rho_i,z) = p_i (z) , \quad 0<z<1, \quad i=0,1,\\[2mm]
\llbracket  w(r,\delta) \rrbracket  = \chi (r) , \quad \alpha < r < \beta,\\[2mm]
\left\llbracket  \displaystyle\frac{\partial w}{\partial z}(r,\delta ) \right\rrbracket  = 0 , \quad \alpha < r < \beta,
\end{array} \right.
\end{equation}
where
$$
\gamma = \frac{a_1 - a_0}{h} , \quad \Psi^2 = \frac{\omega^2 \Omega^2 (a_1 - a_0)^2}{c^2} , \quad \rho_i = \frac{a_i}{a_1 - a_0} , \quad i = 0,1,
$$
$$
\alpha = \frac{c_0}{a_1 - a_0} , \quad \beta = \frac{c_1}{a_1 - a_0} , \quad \delta = \frac{d}{h}.
$$

\section{Displacement within the cylinder}\label{Sect:Displacement}

\subsection{The form of the displacement}

To obtain the displacement function satisfying \eqref{Problem_Formulation_1}, we begin by reducing this to a 1D problem utilizing the finite Fourier sin transform with respect to the variable $z$ (for details on this transform, see e.g. \cite{Debnath2015}). The transformed displacement takes the following form
\begin{equation}\label{transformed_displacement_1}
w_k(r) = \int_0^1 w(r,z) \sin(\lambda_k z ) \, dz , \quad \lambda_k = \frac{\pi}{2}(2k - 1), \quad k=1,2,3,\hdots ,
\end{equation}
with $w(r,z)$ recovered utilizing the associated inverse
\begin{equation} \label{inverse_transform_1}
w(r,z) = 2 \sum_{k=1}^\infty w_k (r) \sin (\lambda_k z ) , \quad \lambda_k = \frac{\pi}{2}(2k - 1).
\end{equation}

Under transformation \eqref{transformed_displacement_1}, the boundary value problem for the displacement \eqref{Problem_Formulation_1} becomes
\begin{equation} \label{Transformed_Problem_1}
\left\{
\begin{array}{l}
\displaystyle\frac{d}{dr}\left( r \displaystyle\frac{d w_k}{dr}\right) - \left( \displaystyle\frac{1}{r} + \mu_k^2 r\right) w_k = \gamma^2 \lambda_k r \cos\left(\lambda_k \delta \right) \chi (r) , \quad \rho_0 < r< \rho_1 , \quad k=1,2,3,\hdots , \\[2mm]
\rho_i \displaystyle\frac{d w_k}{dr}(\rho_i) - w_k (\rho_i) = p_{ik} , \quad i=0,1, \quad k=1,2,3,\hdots ,
\end{array}
\right.
\end{equation}
where $\mu_k = \sqrt{\gamma^2 \lambda_k^2 - \Psi^2}$, while $p_{ik}$ is the transformed tangential loading.

The general solution to the transformed problem \eqref{Transformed_Problem_1} can be expressed as
\begin{equation}\label{transformed_displacement_2}
w_k (r) = A_k I_1 \left(r \mu_k \right) + B_k K_1 \left( r \mu_k \right) + \gamma^2 \lambda_k \cos\left( \lambda_k \delta \right)\int_\alpha^\beta \Phi_k (r,\eta) \eta \chi(\eta) \, d\eta, \quad k=1,2,3,\hdots ,
\end{equation}
where $A_k$, $B_k$ are unknown constants to be obtained from the boundary conditions \eqref{Transformed_Problem_1}$_2$, while
$$
\Phi_k (r,\eta)  =  -\int_0^\infty \frac{x J_1 \left( rx \right) J_1 \left(\eta x \right) }{x^2 + \mu_k^2} \, dx =
 - \left\{ 
 \begin{array}{l} 
  K_1 \left( r \mu_k \right) I_1 \left(\eta \mu_k \right) , \quad \eta < r,\\
  I_1 \left( r \mu_k \right) K_1 \left( \eta \mu_k \right) , \quad \eta > r ,
 \end{array}
 \right.
$$
with $J_n (.)$ denoting the Bessel function, and $I_n (.)$, $K_n (.)$ being the modified Bessel functions of the first and second kind respectively (see e.g. \cite{Abramowitz1972}).

Applying the inverse Fourier sin tranform \eqref{inverse_transform_1}, the normalized displacement experienced by the cylinder immediately follows 
\begin{equation} \label{Normalized_Displacement}
\begin{aligned}
w(r,z) &= 2 \sum_{k=1}^\infty \left[ F_{1k}(r) p_{1k} - F_{0k} p_{0k} \right] \sin\left(\lambda_k z\right) - 2\gamma^2 \int_\alpha^\beta \left[ \sum_{k=1}^\infty \lambda_k \cos\left(\lambda_k \delta\right) \sin\left(\lambda_k z\right) N_k \left(r,\eta\right)\right]\eta \chi (\eta) \, d\eta \\
&\quad + 2\gamma^2 \int_\alpha^\beta \left[ \sum_{k=1}^\infty \lambda_k \cos\left(\lambda_k \delta\right) \sin\left(\lambda_k z\right) \Phi_k \left(r,\eta\right)\right]\eta \chi (\eta) \, d\eta,
\end{aligned}
\end{equation}
where
$$
F_{ik}(r) = \frac{1}{\Delta_k}\left[ K_2 \left(\rho_i \mu_k \right) I_1 \left(r\mu_k \right) + I_2 \left(\rho_i \mu_k \right) K_1 \left( r \mu_k \right) \right] , \quad i = 0,1, \quad k = 1,2,3,\hdots,
$$
$$
\begin{aligned}
N_k (r, \eta ) &= \frac{1}{\Delta_k} \left[ K_2 (\rho_1 \mu_k ) I_1 (r  \mu_k) \left\{ I_2 (\rho_0 \mu_k ) K_1 (\eta \mu_k ) + K_2 (\rho_0 \mu_k) I_1 (\eta \mu_k) \right\} \right. \\
&\quad \left. + I_2 (\rho_0 \mu_k ) K_1 (r\mu_k ) \left\{ K_2 (\rho_1 \mu_k)I_1 (\eta\mu_k ) + I_2 (\rho_1 \mu_k) K_1 (\eta \mu_k ) \right\} \right] , \quad k = 1,2,3,\hdots ,
\end{aligned}
$$
with 
$$
\Delta_k = I_2 \left(\rho_1 \mu_k \right) K_2 \left(\rho_0 \mu_k \right) - K_2 \left(\rho_1 \mu_k \right) I_2 \left(\rho_0 \mu_k \right), \quad k = 1,2,3,\hdots.
$$

While the form of the displacement has now been obtained, it is in terms of the still unknown jump function $\chi( r )$, which must be computed.

\subsection{The jump of the displacement}\label{Sect:Traction}

\subsubsection{The singular integral equation}

The unknown function $\chi ( r )$ is obtained from the remaining boundary condition $\left\llbracket \displaystyle\frac{dw}{dz}(r,\delta)\right\rrbracket =0$ \eqref{Problem_Formulation_1}. In order to differentiate \eqref{Normalized_Displacement}, weakly convergent parts are summed up. The condition is then expressed in the form of a singular integral equation. The full details on the derivation of this equation are provided in \cite{VAYSFELD2017526} (see also references therein), and as such only a summary is provided here.

We can express the condition in terms of an unknown function, $t$, dependent on the displacement jump, $\chi$, as
\begin{equation} \label{Function_t}
 t(\xi ) = \exp\left( \frac{1+\xi}{2\nu}\right) \chi^* \left(\alpha \exp\left[ \frac{1+\xi}{\nu}\right] \right) , \quad \chi^* (r) = \frac{d}{dr}\left[ r \chi(r)\right], \quad \xi = \frac{2\ln\left(r / \alpha\right)}{\ln\left(\beta / \alpha\right)} - 1, \quad \alpha < r < \beta,
\end{equation}
with $\nu = 2\left[ \ln \left({\beta} / {\alpha}\right)\right]^{-1}$. The singular integral equation then takes the form
\begin{equation} \label{Traction_1}
 \int_{-1}^1 \left[ - \ln\left|s-\xi\right| + l^* (s-\xi ) + L(s,\xi)\right] t(\xi ) \, d\xi = T(s) - C \cdot h(s), \quad -1 < s < 1,
\end{equation}
where
$$
L(s,\xi) = \pi \alpha e^{\frac{2+s+\xi}{2\nu}} R^* \left( \alpha e^{\frac{1+s}{\nu}}, \alpha e^{\frac{1+\xi}{\nu}}\right) , \quad T(s) = \pi \nu e^{\frac{1+s}{2\nu}} M^* \left( \alpha e^{\frac{1+s}{\nu}} \right),
$$
$$
h(s) = \pi \nu e^{\frac{1+s}{2\nu}} , \quad l(x) = \frac{1}{\cosh\left(\frac{x}{2\nu}\right)} K \left( \frac{1}{\cosh\left(\frac{x}{2\nu}\right)} \right) = -\ln\left|x\right| + l^*(x) , \quad \lim_{x\to 0} l^* (x) = \ln\left( 8\nu \right) , 
$$
with
$$
\begin{aligned}
R^* (r, \eta ) &= \int_0^\infty \frac{J_1 (rx) J_1 (\eta x )}{x} \left[ \frac{\sqrt{x^2 - \Psi^2}}{\cosh\left(\frac{1}{\gamma}\sqrt{x^2 - \Psi^2}\right)} \sinh \left( \frac{1}{\gamma}\sqrt{x^2 - \Psi^2}(1-2\delta )\right) + \sqrt{x^2 - \Psi^2} \tanh\left( \frac{1}{\gamma}\sqrt{x^2 - \Psi^2}\right) - x\right] \, dx \\
& \quad + 4\gamma^3 \sum_{k=1}^\infty \lambda_k^2 \cos\left(\lambda_k \delta\right)N_k^* (r,\eta) ,
\end{aligned}
$$
$$
M^* (r) = -4\gamma \sum_{k=1}^\infty \left[ F_{1k}^* (r) p_1k - F_{0k}^* (r) p_{0k} \right] \lambda_k \cos\left(\lambda_k \delta \right) , 
$$
and
$$
F_{ik}^* (r) = \frac{1}{\Delta_k} \left[ K_2 \left( \rho_i \mu_k \right) I_0 \left( r \mu_k \right) - I_2 \left( \rho_i \mu_k \right) K_0 \left( r \mu_k \right) \right] , \quad i = 0,1,
$$
$$
\begin{aligned}
N_k^* (r,\eta) &= -\frac{1}{\mu_k^2 \Delta_k} \left[ K_2 \left( \rho_1 \mu_k \right) I_1 \left(r\mu_k \right) \left\{ I_2 \left( \rho_0 \mu_k \right) K_1 \left( \eta \mu_k \right) + K_2 \left( \rho_0 \mu_k \right) I_1 \left( \eta \mu_k \right) \right\} \right. \\
&\left. \quad + I_2 \left(\rho_0 \mu_k \right) K_1 \left( r \mu_k \right) \left\{ K_2 \left(\rho_1 \mu_k \right) I_1 \left(\eta \mu_k \right) + I_2 \left( \rho_1 \mu_k \right) K_1 \left(\eta \mu_k \right) \right\} \right],
\end{aligned}
$$
with $K(.)$ being the complete elliptic integral of the first kind (see e.g. \cite{Abramowitz1972}).

The problem of obtaining the jump function satisfying the boundary condition now consists of finding the solution $t(\xi)$ of the singular integral equation \eqref{Traction_1}, and inverting expression \eqref{Function_t} to obtain $\chi (r)$.

\subsubsection{Form of the function $t$}

The solution to the singular integral equation \eqref{Traction_1} is sought utilizing the method of orthogonal polynomials first proposed in \cite{Popov1982}. Accordingly, we seek the function $t(\xi)$ in the form
\begin{equation}\label{traction_form}
t(\xi) = \frac{1}{\sqrt{1-\xi^2}} \sum_{n=0}^\infty \left( t_n + C \cdot t_n^C \right) T_n (\xi),
\end{equation}
where $t_n$, $C$, $t_n^C$, with $n=0,1,2,\hdots$, are unknown constants, while $T_n (\xi)$ are Chebyshev polynomials of the first kind (see e.g. \cite{Abramowitz1972}).

Before inserting expression \eqref{traction_form} into the singular integral equation \eqref{Traction_1}, we observe the following spectral correspondence
\begin{equation}\label{spectral_correspondence}
\int_{-1}^1  \ln\left| s-\xi \right| \frac{T_n (\xi)}{\sqrt{1-\xi^2}} \, d\xi = - \sigma_n T_n (s), \quad \sigma_n = \begin{cases} \pi \ln(2) , & n=0,\\ \displaystyle\frac{\pi}{n}, & n=1,2,\hdots .\end{cases}
\end{equation}
Therefore, multiplying \eqref{Traction_1} through by $T_n(s) / \sqrt{1-s^2}$ and integrating over $s\in [-1,1]$, we obtain
\begin{equation} \label{modified_traction}
\tilde{t}_m + C\cdot \tilde{t}^C_m + \sum_{n=0}^\infty \left( \tilde{t}_n + C\cdot \tilde{t}_n^C \right) A_{mn} = b_m + C\cdot b_m^C, \quad m=0,1,2,\hdots,
\end{equation}
where
$$
\tilde{t}_m = \sqrt{\sigma_m \gamma_m} t_n , \quad \tilde{t}_m^C = \sqrt{\sigma_m \gamma_m} \sigma_m^C, \quad b_m^C = \frac{\pi \nu}{\sqrt{\sigma_m \gamma_m}} e^{\frac{1}{2\nu}} J_m \left(\frac{1}{2\nu}\right) , \quad \gamma_m = \begin{cases} \pi, & m=0, \\ \displaystyle\frac{\pi}{2} , & m=1,2,\hdots , \end{cases}
$$
$$
A_{mn} = \frac{1}{\sqrt{\sigma_m \gamma_m \sigma_n \gamma_n}}  \int_{-1}^1 \frac{T_m (s)}{\sqrt{1-s^2}} \, ds  \int_{-1}^1 \left[ l^* (s-\xi ) + L(s,\xi) \right] \frac{T_n (\xi)}{\sqrt{1-\xi^2}} \, d\xi , \quad b_m = \frac{1}{\sqrt{\sigma_m \gamma_m}} \int_{-1}^1 \frac{T_m (s)}{\sqrt{1-s^2}} \, ds .
$$
The only unknowns to be solved for are the constant $C$ and the constants $t_n$, $t_n^C$, $n=0,1,2,\hdots$. 

The value of the constant $C$ follows from expanding the singular integral equation \eqref{Traction_1}, and noting that the jump function at the crack surfaces $\chi (\alpha) = \chi (\beta) = 0$ (see \cite{VAYSFELD2017526} for details)
\begin{equation} \label{Const_C}
C = - \sum_{n=0}^\infty \frac{\tilde{t}_n}{\sqrt{\sigma_n \gamma_n}} I_n \left(\frac{1}{2\nu}\right) \left[ \sum_{m=0}^\infty \frac{\tilde{t}_m^C}{\sqrt{\sigma_m \gamma_m}} I_m \left(\frac{1}{2\nu}\right) \right]^{-1} .
\end{equation}
Consequently, the constants $t_n$, $t_n^C$, $n=0,1,2,\hdots$, now follow immediately from \eqref{modified_traction}. This allows the displacement jump $\chi (r)$ to be obtained, yielding a full description of the displacement within the cylinder.

\section{The stress intensity factor}\label{Sect:SIFs}

We investigate the stress intensity factor experienced on the opposing crack faces, in order to determine the crack behaviour and potential for extension (see e.g. \cite{PICCOLROAZ2021103487}). The normalization of this parameter is taken in line with \eqref{Normalization} as
\begin{equation} \label{Normalization_K_III}
K_{III} = \frac{\tilde{K}_{III} \sqrt{a_1 - a_0}}{F},
\end{equation}
for $\tilde{K}_{III}$ the (dimensional) mode-III stress intensity factor, and $F$ the maximal applied load.

The dimensionless stress intensity factors are then given by
$$ 
K_{III}^- = \lim_{r\to c_0^-} \sqrt{2\pi (c_0 - r)} \tau_{\phi z} , \quad K_{III}^+ = \lim_{r\to c_1^+} \sqrt{2\pi (c_1 - r)} \tau_{\phi z} .
$$ 
Noting the form of the displacement \eqref{traction_form}, the tangential stress $\tau_{\phi z}$ follows immediately from the stress - displacement relations in cylindrical coordinates. The stress intensity factor is thus obtained as (for full details of evaluating this limit, see \cite{VAYSFELD2017526})
\begin{equation} \label{SIFs}
K_{III}^- = \frac{G\sqrt{\pi}}{2\sqrt{c_0 \nu}} \sum_{n=0}^\infty (-1)^n \left( t_n + C\cdot t_n^C \right) , \quad K_{III}^+ = \frac{G\sqrt{\pi}}{2\sqrt{c_1 \nu}} \sum_{n=0}^\infty \left( t_n + C\cdot t_n^C \right) .
\end{equation} 

\section{Results for a crack in a rotating cylinder}\label{Sect:NumInvest}

\subsection{Numerical scheme and simulation parameters}\label{Sect:Parameters}

The system of equations is solved using an iterative scheme in a Python environment. The jump function is obtained by solving \eqref{modified_traction} - \eqref{Const_C}, inserting these constants into \eqref{traction_form} to yield the function $t(\xi)$, and then solving the inverse problem \eqref{Function_t} to obtain $\chi (r)$. The displacement, $w(r,z)$, then follows immediately from \eqref{Normalized_Displacement} (noting the boundary conditions), while the stress intensity factor on the crack surfaces, $K_{III}^{+/-}$, is obtained using \eqref{SIFs}. 

\begin{table}[t]
 \centering
 \begin{tabular}{c | c | c | c | c | c}
 \hline \hline
 $a_0$ [m] & $a_1$ [m] & $c_1$ [m] & $h$ [m] & $\Omega$ [Hz] & $F$ [Pa] \\
 \hline
 $5.5 \times 10^{-3}$ & $7.5\times 10^{-3}$ & $7 \times 10^{-3}$ & $5 \times 10^{-3}$ & $150$ & $3\times 10^{5}$ \\
 \hline \hline
\end{tabular}
\put(-350,30) {{\bf (a)}}

\vspace{4mm}

\begin{tabular}{r | c | c }
\hline \hline 
Configuration & $d$ [m] & $c_0$ [m] \\
\hline
 Symmetric & $h/2$ & $ a_0 + (a_1 - a_0)/4$ \\
 \hline
 Edge & $h/10$ & $a_0 + (a_1 - a_0)/15$ \\
\hline \hline 
\end{tabular}
\put(-297,30) {{\bf (b)}}
\caption{{\bf (a)} The value of constants used in simulations. {\bf (b)} The four configurations of the crack location used in simulations (symmetric - symmetric, symmetric - edge, edge - symmetric and edge - edge). Here, the edge cases represent the minimal distance to the bottom of the cylinder and the inner surface for $d$ and $c_0$ respectively.}
\label{Table_Constants}
\end{table}

In simulations, the constants defining the problem geometry and material constants are taken for a small steel cylinder, unless otherwise stated. The values used for the problem geometry are given in Table.~\ref{Table_Constants}a. To examine the influence of the crack location, we vary the height and inner radius of the fracture in four configurations, as outlined in Table.~\ref{Table_Constants}b.

For the prescribed loading, we assume that the cylinder is not experiencing loading on the outer surface ($P_1 (z)=0$), with all loading occurring on the inner surface ($P_0(z)\neq 0$). For simplicity, we will stick to power-law loading, for example constant loading $P_0 (z) = F$ (with $F$, the maximal applied load, given in Table.~\ref{Table_Constants}a), linear loading $P_0 (z) = Fz$, etc. Simulations considering piece-wise loading were also conducted, such that only a portion of the cylinder inner surface experienced loading, and the results were inline with those presented in the remainder of this section (for this reason, they are not included). 

\subsection{The dependence of crack extension on process parameters}\label{Sect:CrackGrowth}

The first immediate concern when considering a crack within the rotating cylinder is the direct damage it may cause. This can primarily be considered in terms of the fracture extension, both in the case of a single event (fast fracture) and over multiple cycles (fatigue crack). Towards this end, we consider the normalized stress intensity factor (SIF) experienced on the crack surfaces \eqref{SIFs}, and their dependence on the fracture location. 

\subsubsection{Quasi-static fracture growth}

\begin{figure}[h!]
 \centering
 \includegraphics[width=0.45\textwidth]{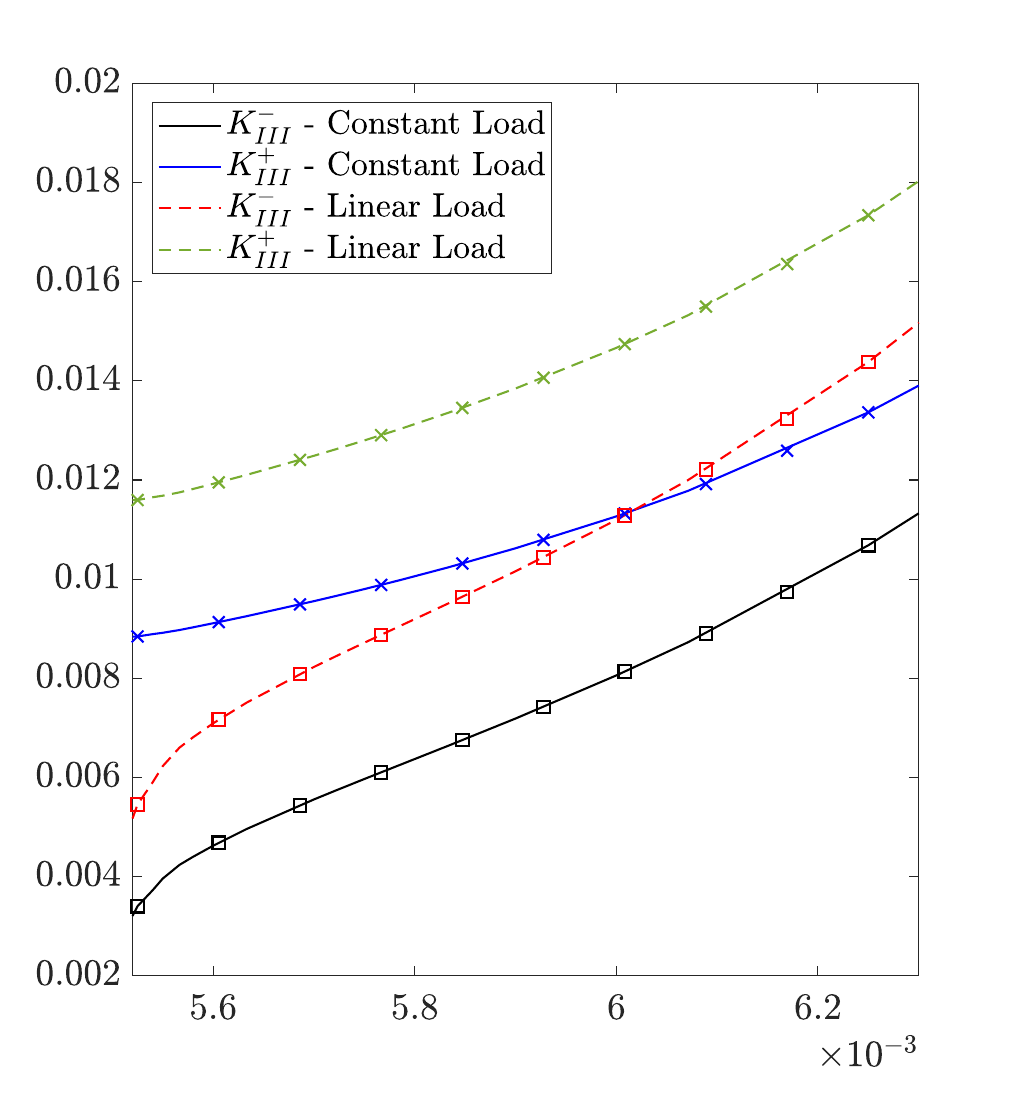}
  \put(-105,0) {$\alpha$}
 \put(-225,115) {$K_{III}$}
 \put(-225,215) {{\bf (a)}}
 \put(-130,215) {{\it Edge crack}}
 \hspace{4mm} 
 \includegraphics[width=0.45\textwidth]{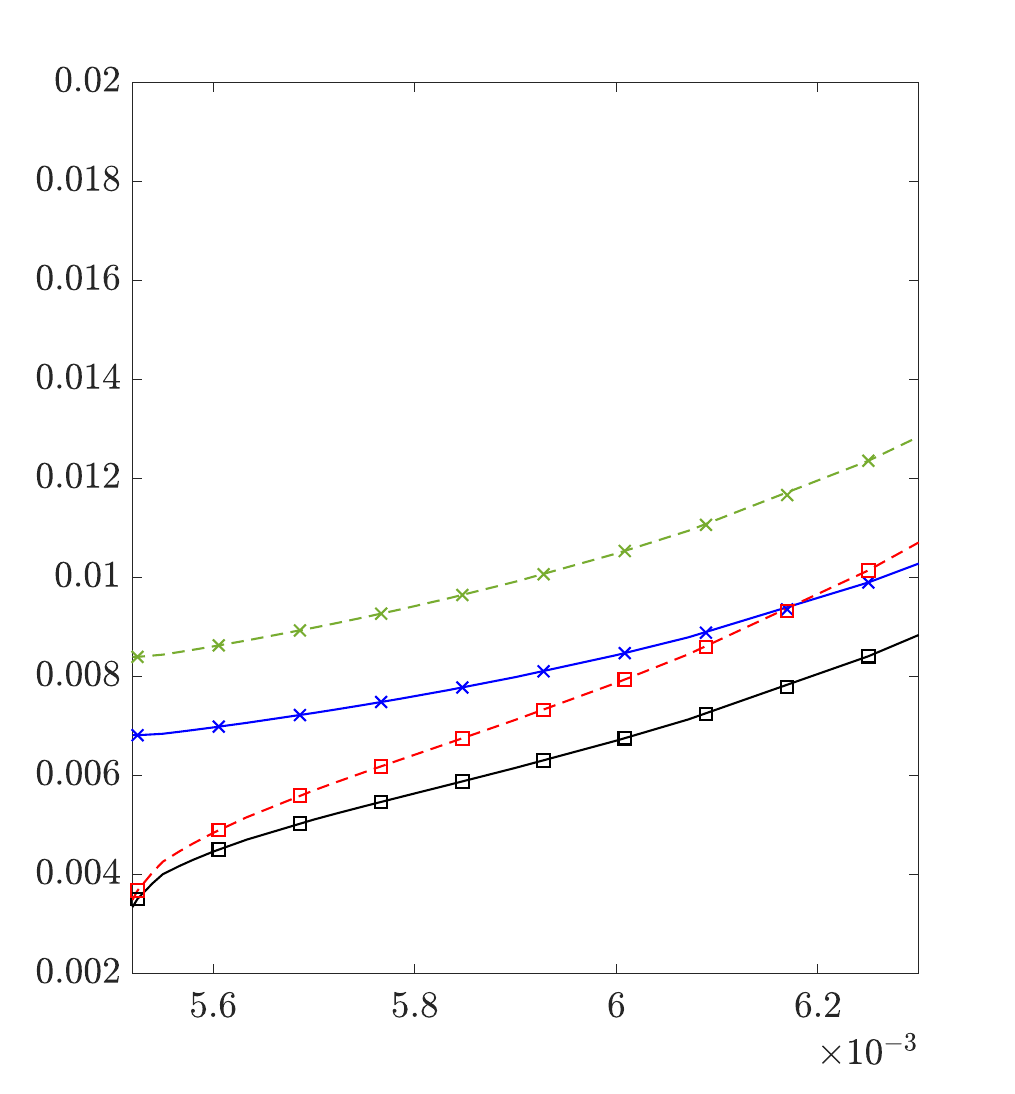}
  \put(-105,0) {$\alpha$}
 \put(-225,115) {$K_{III}$}
 \put(-225,215) {{\bf (b)}}
 \put(-140,215) {{\it Symmetric crack}}

\vspace*{4mm}

  \includegraphics[width=0.45\textwidth]{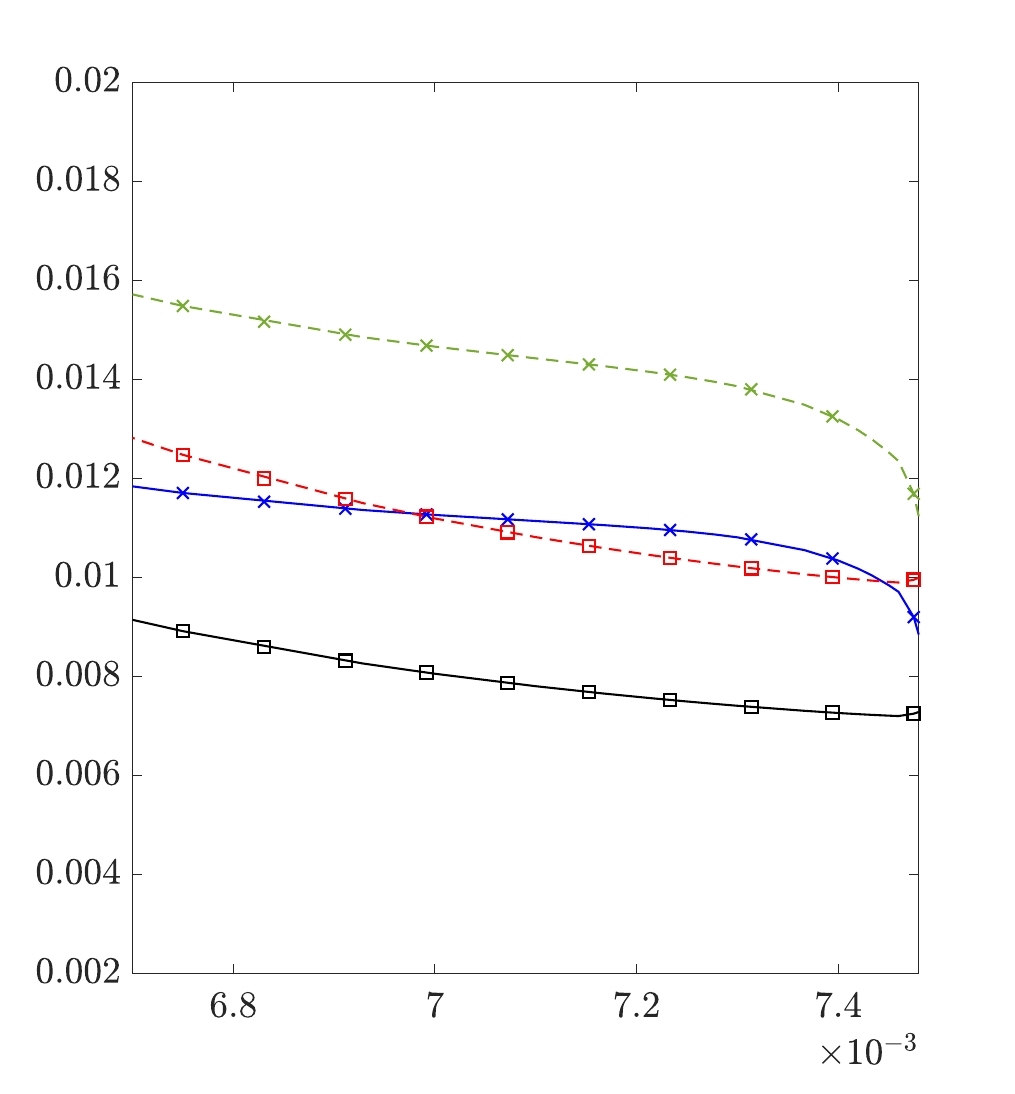}
 \put(-105,0) {$\beta$}
 \put(-225,115) {$K_{III}$}
 \put(-225,215) {{\bf (c)}}
 \put(-130,215) {{\it Edge crack}}
 \hspace{4mm} 
 \includegraphics[width=0.45\textwidth]{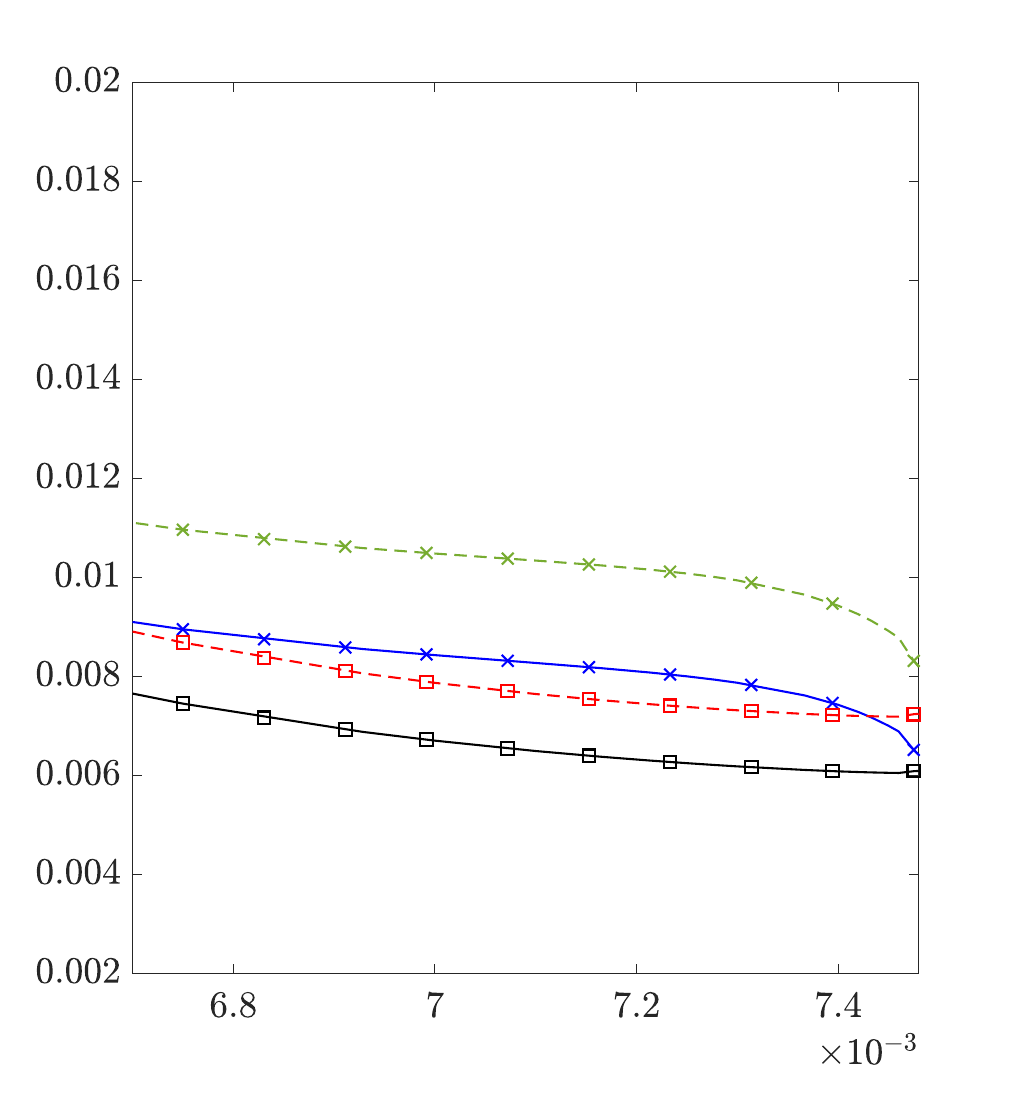}
 \put(-105,0) {$\beta$}
 \put(-225,115) {$K_{III}$}
 \put(-225,215) {{\bf (d)}}
 \put(-140,215) {{\it Symmetric crack}}
 \caption{Dependence of the dimensionless stress intensity factor on the inner ($K_{III}^{-}$, square markers) and outer ($K_{III}^+$, cross markers) fracture surfaces on the normalized crack {\bf (a)}, {\bf (b)} inner radius $\alpha$, {\bf (c)}, {\bf (d)} outer radius $\beta$. Results are shown for constant loading (unbroken lines) and linear loading (dashed lines), in the case of an {\bf (a)}, {\bf (c)} edge crack, {\bf (b)}, {\bf (d)} symmetric crack (see Table.~\ref{Table_Constants}b). All remaining parameters for the cylinder and crack geometry are taken as stated in Table.~\ref{Table_Constants}a.}
 \label{Fig:c0}
\end{figure}

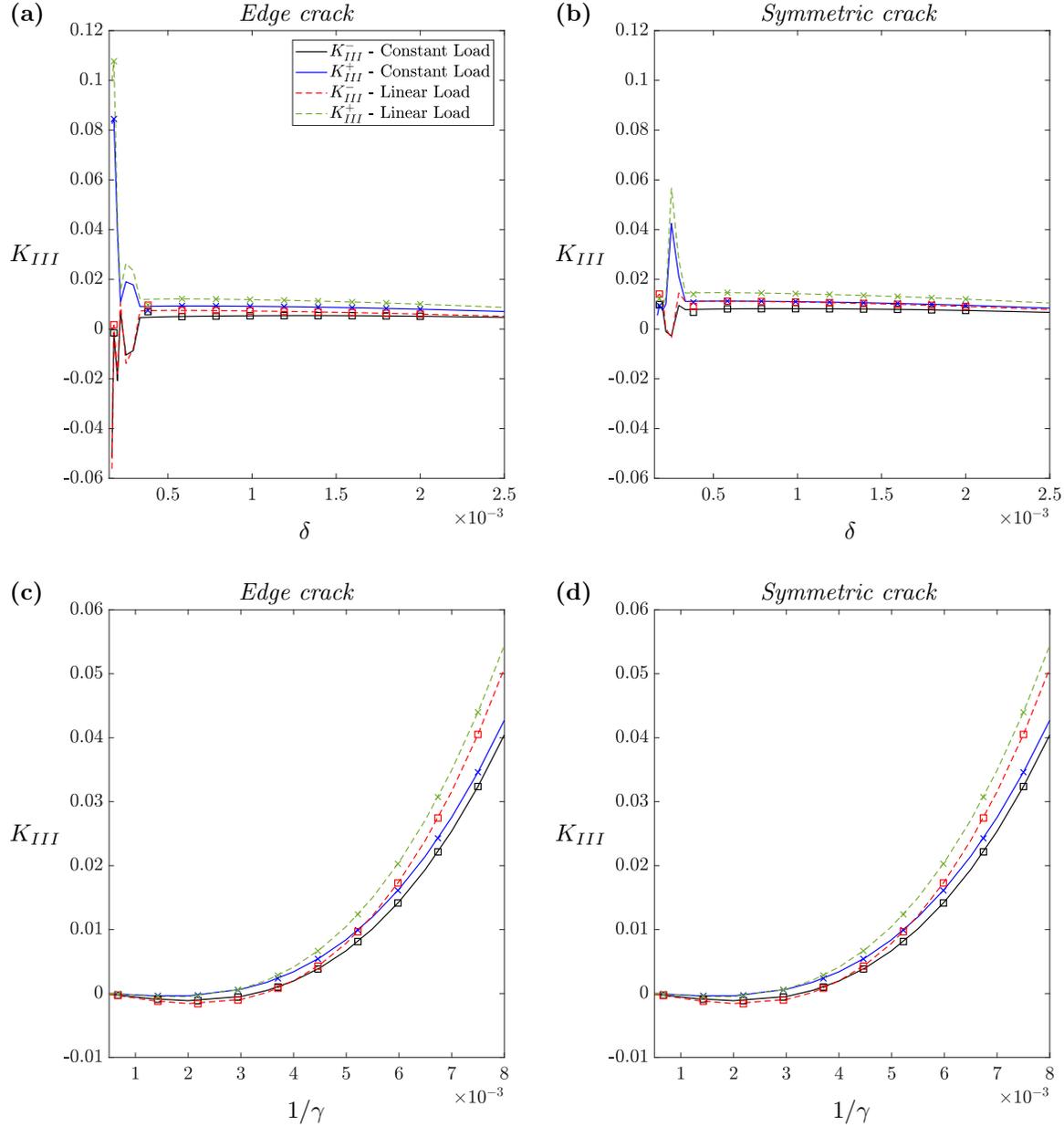
\begin{figure}[t!]
 \centering
 \includegraphics[width=0.45\textwidth]{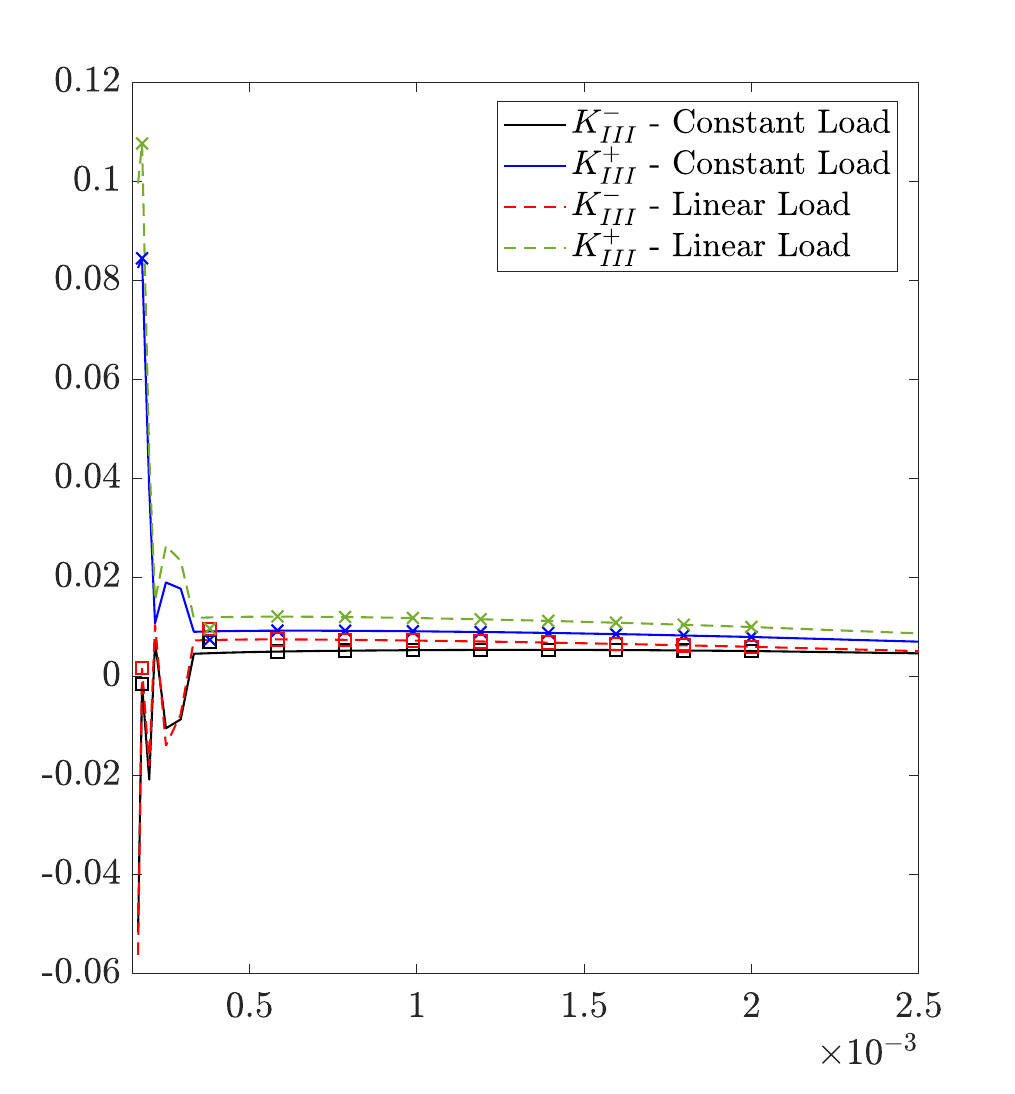}
  \put(-105,0) {$\delta$}
 \put(-225,115) {$K_{III}$}
 \put(-225,215) {{\bf (a)}}
 \put(-130,215) {{\it Edge crack}}
 \hspace{4mm} 
 \includegraphics[width=0.45\textwidth]{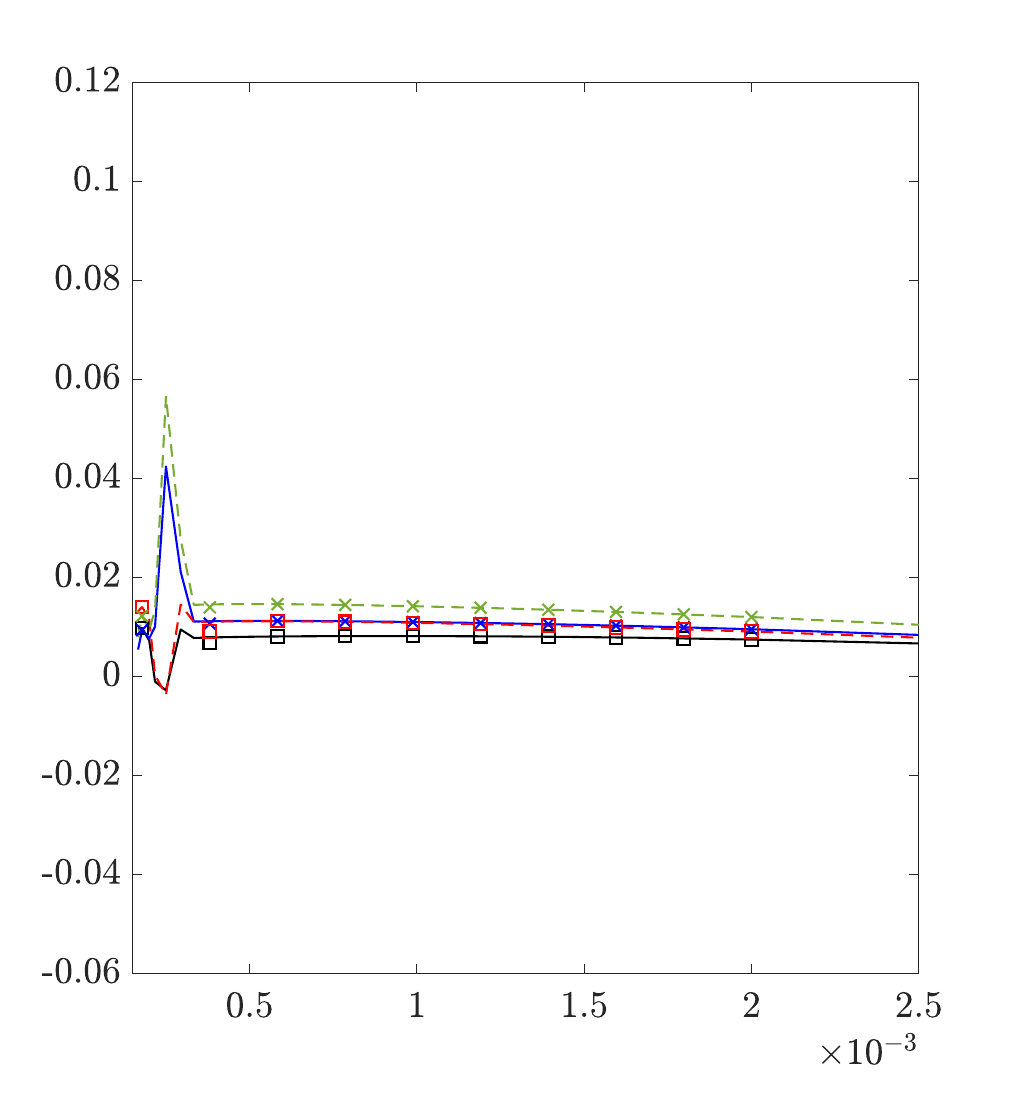}
  \put(-105,0) {$\delta$}
 \put(-225,115) {$K_{III}$}
 \put(-225,215) {{\bf (b)}}
 \put(-140,215) {{\it Symmetric crack}}

\vspace*{4mm}

  \includegraphics[width=0.45\textwidth]{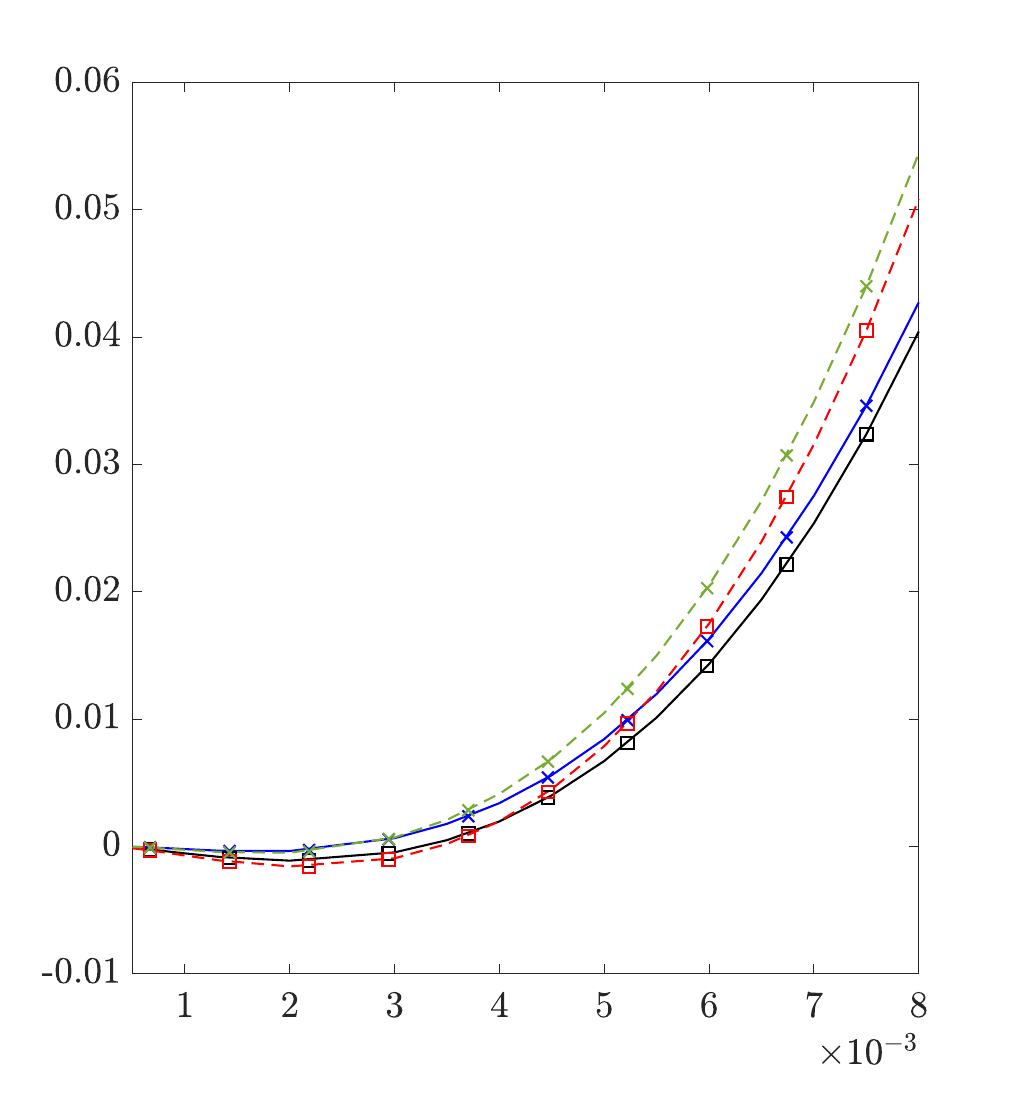}
 \put(-110,0) {$1 / \gamma$}
 \put(-225,115) {$K_{III}$}
 \put(-225,215) {{\bf (c)}}
 \put(-130,215) {{\it Edge crack}}
 \hspace{4mm} 
 \includegraphics[width=0.45\textwidth]{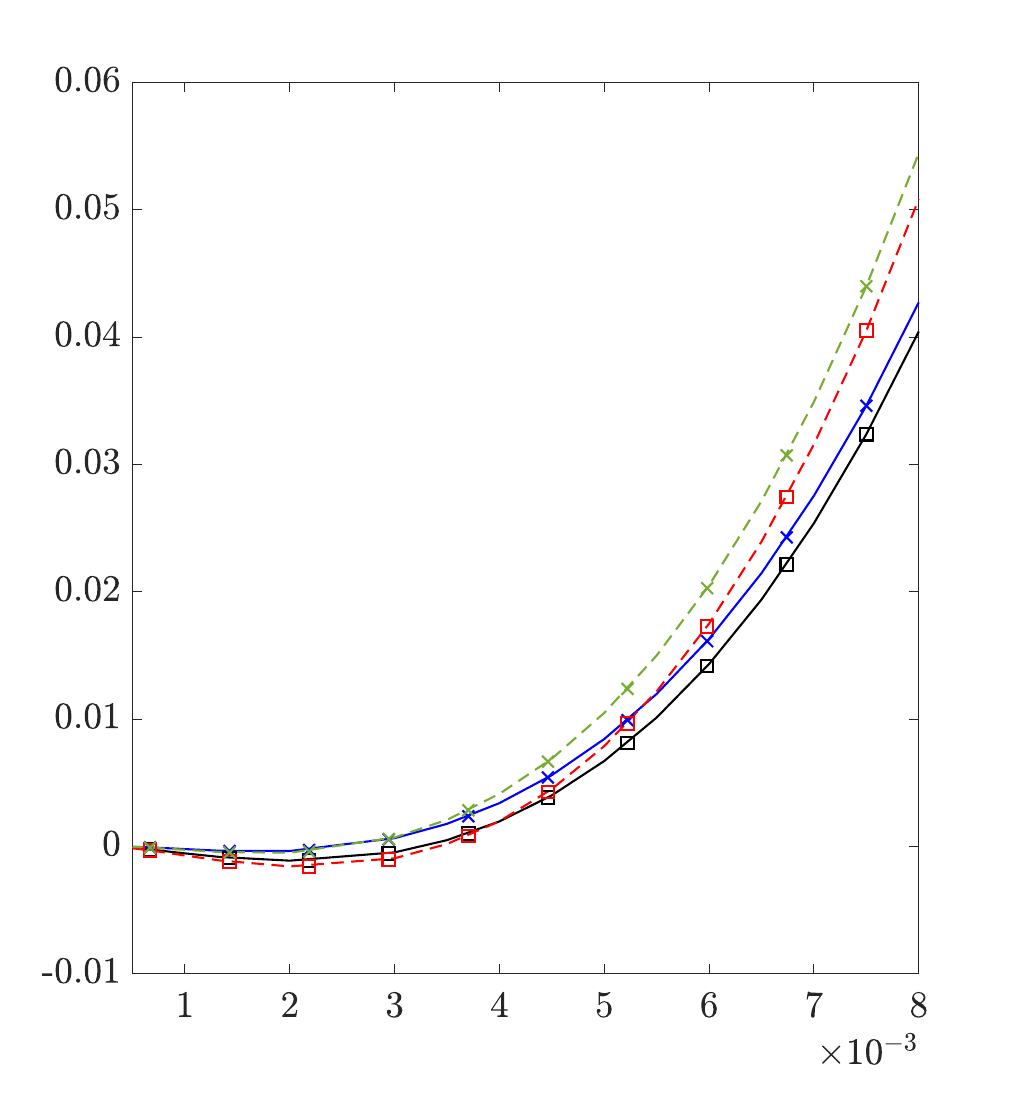}
 \put(-110,0) {$1 / \gamma$}
 \put(-225,115) {$K_{III}$}
 \put(-225,215) {{\bf (d)}}
 \put(-140,215) {{\it Symmetric crack}}
 \caption{Dependence of the dimensionless stress intensity factor on the inner ($K_{III}^{-}$, square markers) and outer ($K_{III}^+$, cross markers) fracture surfaces on the normalized {\bf (a)}, {\bf (b)} crack height $\delta$, {\bf (c)}, {\bf (d)} cylinder height $1/\gamma = h/(a_1 - a_0)$. Results are shown for constant loading (unbroken lines) and linear loading (dashed lines), in the case of an {\bf (a)}, {\bf (c)} edge crack, {\bf (b)}, {\bf (d)} symmetric crack (see Table.~\ref{Table_Constants}b). All remaining parameters for the cylinder and crack geometry are taken as stated in Table.~\ref{Table_Constants}a.}
 \label{Fig:h}
\end{figure}

%
Let us first consider the case of a single cycle/loading event. In this instance, the stress intensity factor can be utilized to predict fracture extension, through use of the energy release rate or similar criterion. When interpreting the results, the presented normalization scheme for $K_{III}$ \eqref{Normalization_K_III} will yield a normalized mode-I material toughness for steel of approximately $K_{Ic}=7.45$ (assuming dimensional $\tilde{K}_{Ic}=50 \, \text{MPa} \cdot \sqrt{\text{m}}$).


The influence of the fracture inner and outer surface positions, $c_0$ and $c_1$ respectively, on the SIFs are provided in Fig.~\ref{Fig:c0}. It is clear that for both parameters the impact of the crack is significantly higher when it is located at the edge of the domain (i.e.\ at the bottom of the cylinder) rather than at the center (compare (a), (c) and (b), (d)). The results also show that the normalized stress intensity factor increases with increasing $c_0$, and decrease with decreasing $c_1$, reflecting the fact that a higher the stress intensity factor experienced for a smaller crack surface under identical loading.

The dependence of the stress intensity factor on the fracture height, $d$, and cylinder height, $h$, are provided in Fig.~\ref{Fig:h}. We again observe a larger impact of the crack in the edge cases (when the crack almost touches the inner cylinder surface), although it is far less pronounced. The results for the crack height indicate that the stress intensity factor is largest when the crack is located near the bottom of the cylinder, with the SIFs falling rapidly as $d$ increases. Modifying the cylinder height $h$ has the opposite effect, with $K_{III}^{+/-}$ becoming negative for a sufficiently small cylinder, but increasing rapidly as the cylinder size increases.

Interestingly, numerical investigations by the authors demonstrated that the influence of the stress intensity factor on the frequency of cylinder rotation, $\omega$, is almost negligible. This is likely due to the small cylinder size being considered, which alongside the high wave speed in the medium leads to the term $\Psi$ being negligible in \eqref{Problem_Formulation_1}, thereby eliminating the rotation-induced effects from the formulation. It follows that the impact of the rotational frequency only becomes significant when considering larger structures.

It is clear from the above results that the crack location has a significant impact on the stress intensity factor experienced on the fracture walls. This is not unexpected, but does mean that the crack location must be determined in order to effectively predict the risk of fast fracture within the cylinder. The issue of determining the crack location is considered in more detail in Sect.~\ref{Sect:Detect}.

\subsubsection{Potential applications to fatigue crack risk management}

The presented formulation can easily be utilized to produce estimates regarding fatigue cracking over multiple loading cycles. The can be achieved by using the presented formulation to obtain quasi-static predictions of the maximum and minimum stress intensity factor experienced during each individual loading cycle, and utilizing Paris' law to predict fracture growth during/between cycles (see for example \cite{PUGNO20061333} and references therein). 

This can be utilized in a number of ways. For example, it could be used during the design of cylindrical components to predict an upper bound on the number of loading cycles for a fatigue crack to reach a certain length. This can then be used to inform the maximum permissible time between component inspections. 

Alternatively, if a fracture was detected, the presented formulation could be utilized to provide an upper bound on the remaining `safe' operation time, or a permissible timeline for repair work to take place. Such applications however require the ability to detect the crack within the cylinder, even though it is not visible.

\subsection{On detecting a crack within a rotating cylinder}\label{Sect:Detect}

One of the most important considerations for cracks within a cylinder is the ability to detect them. Being able to determine their location is also important, as this allows local fixes (repairs or strengthening) to be applied. We therefore consider tests which can be applied to a rotating cylinder, in order to determine the crack location.

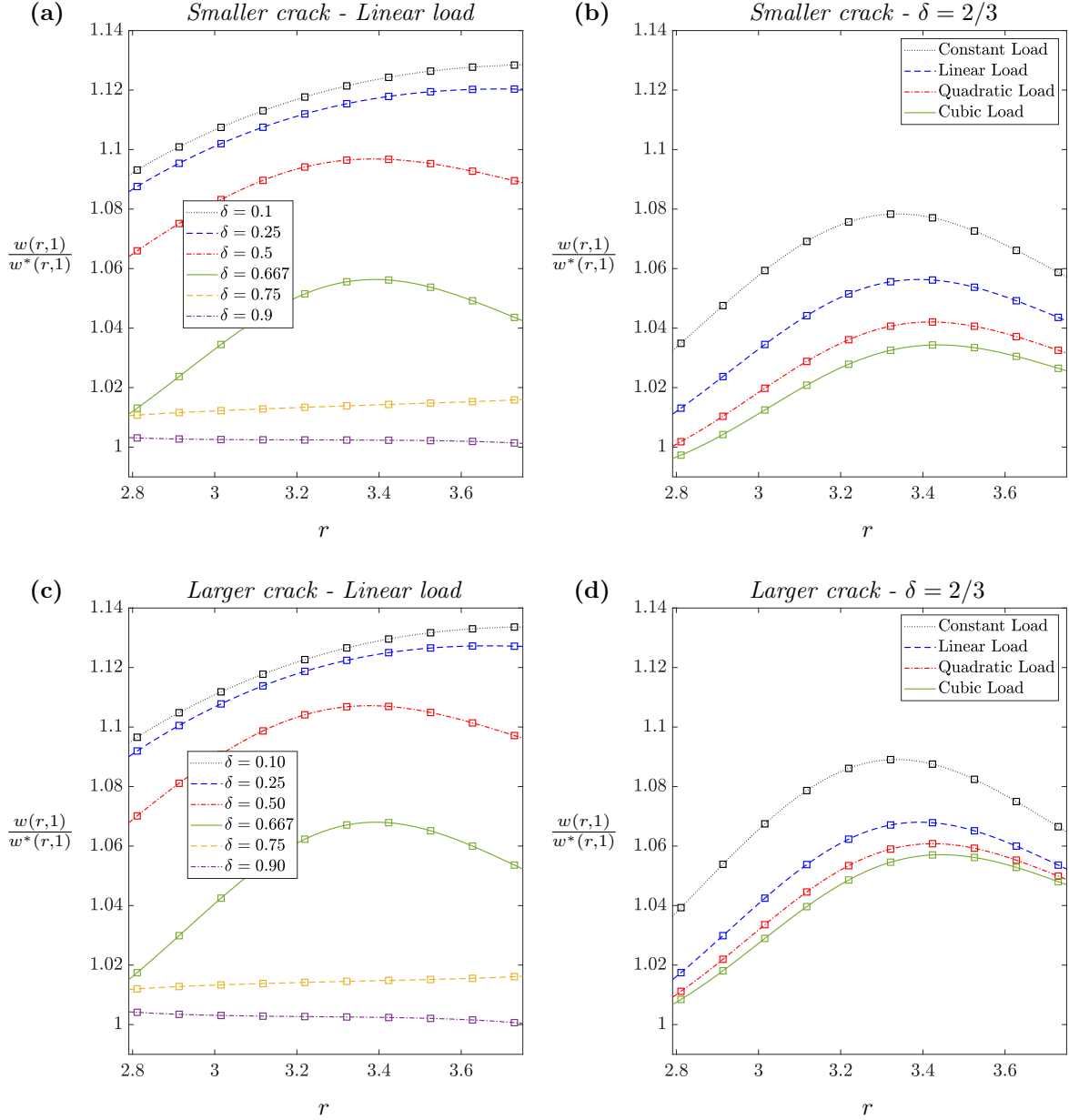
\begin{figure}[t!]
 \centering
  \includegraphics[width=0.45\textwidth]{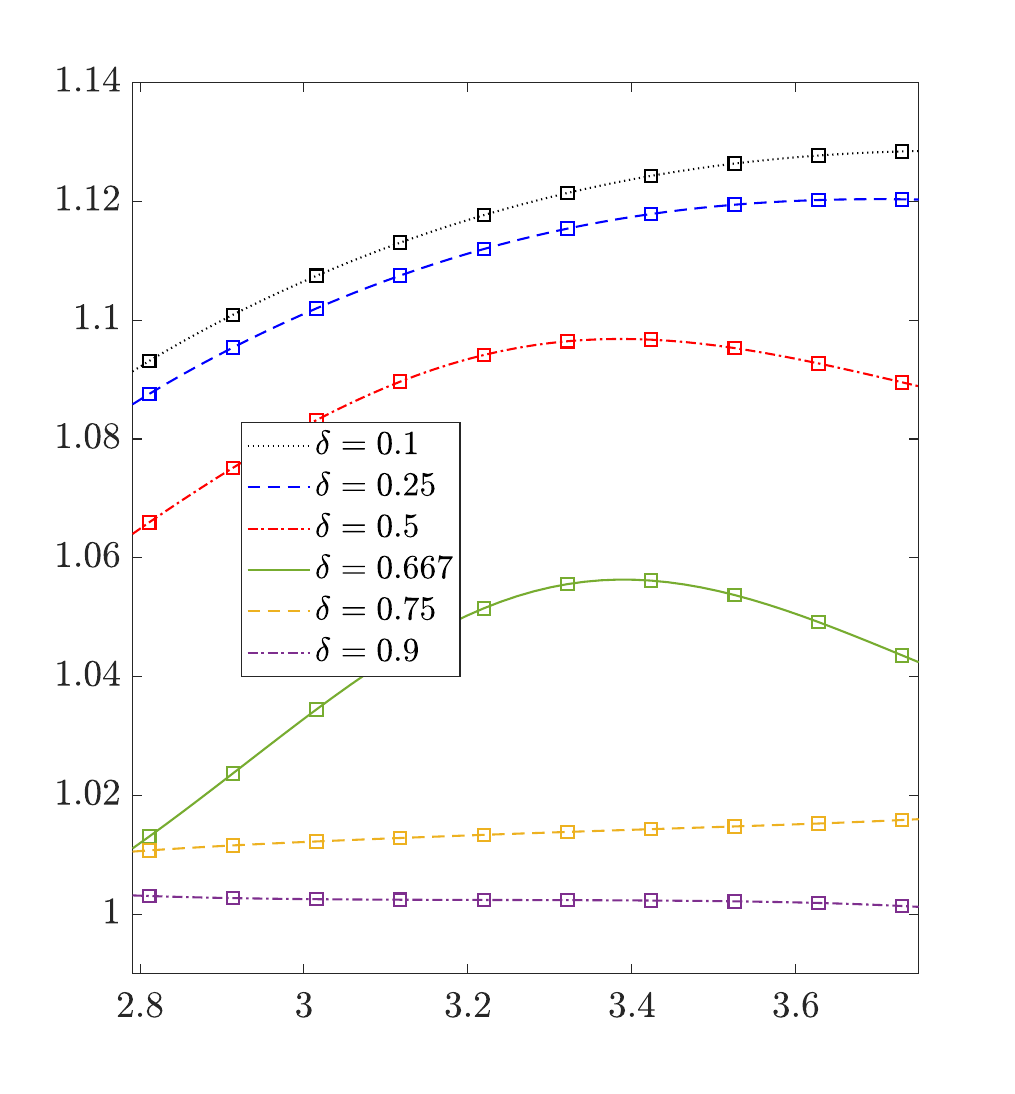}
  \put(-105,0) {$r$}
 \put(-235,115) {$\frac{w(r,1)}{w^*(r,1)}$}
 \put(-225,215) {{\bf (a)}}
 \put(-160,215) {{\it Smaller crack - Linear load}}
 \hspace{4mm} 
 \includegraphics[width=0.45\textwidth]{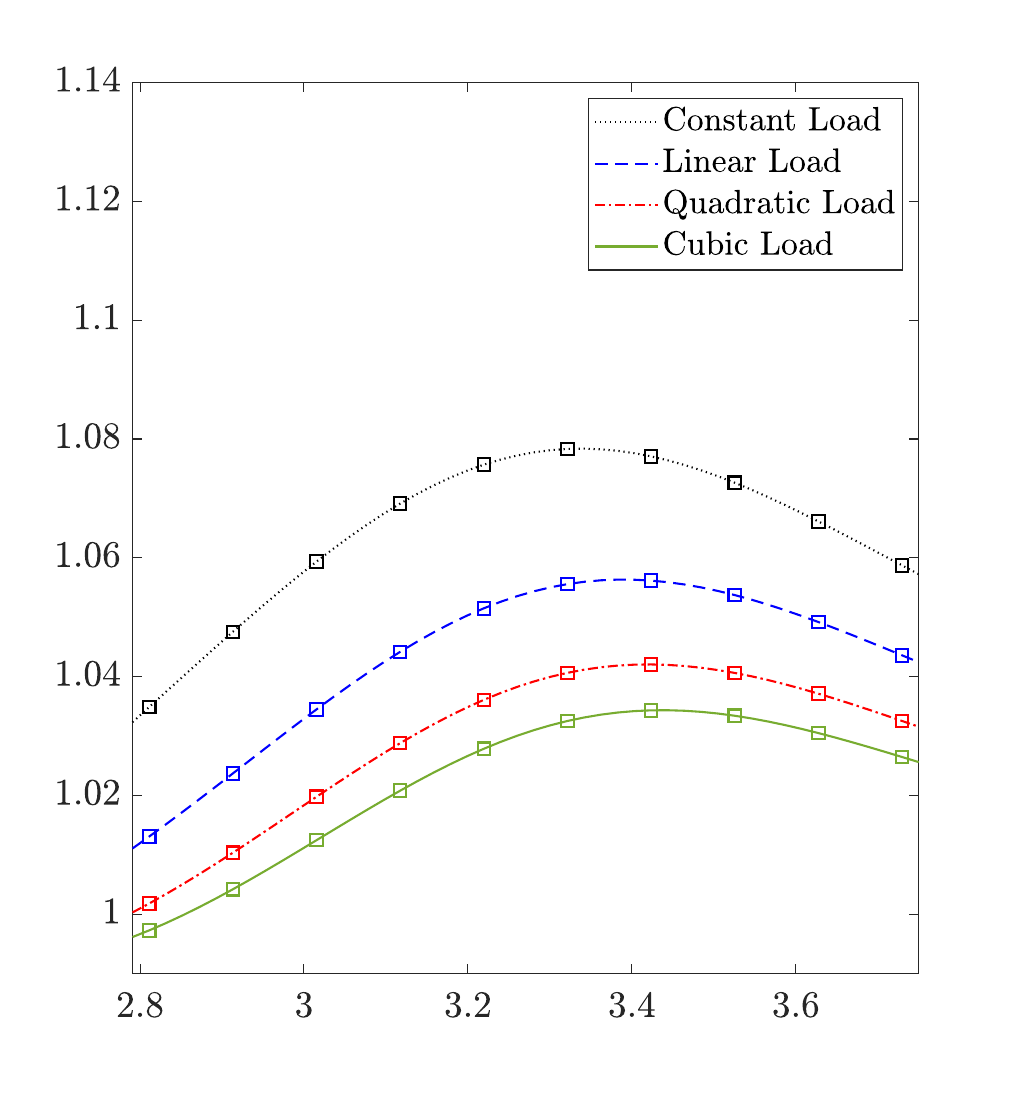}
  \put(-105,0) {$r$}
 \put(-235,115) {$\frac{w(r,1)}{w^*(r,1)}$}
 \put(-225,215) {{\bf (b)}}
 \put(-152,215) {{\it Smaller crack - $\delta = 2/3$}}
 
\vspace*{4mm}

   \includegraphics[width=0.45\textwidth]{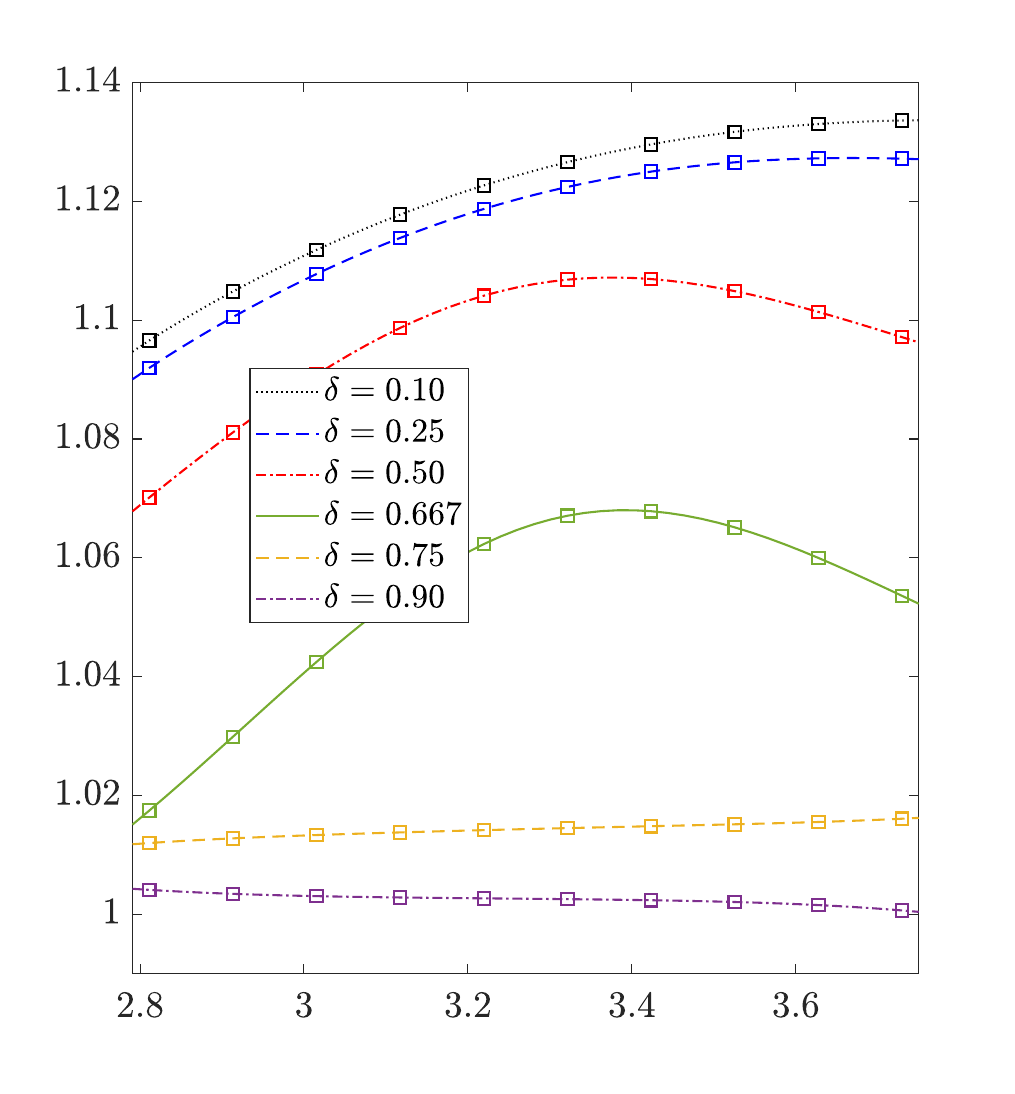}
  \put(-105,0) {$r$}
 \put(-235,115) {$\frac{w(r,1)}{w^*(r,1)}$}
 \put(-225,215) {{\bf (c)}}
 \put(-160,215) {{\it Larger crack - Linear load}}
 \hspace{4mm} 
 \includegraphics[width=0.45\textwidth]{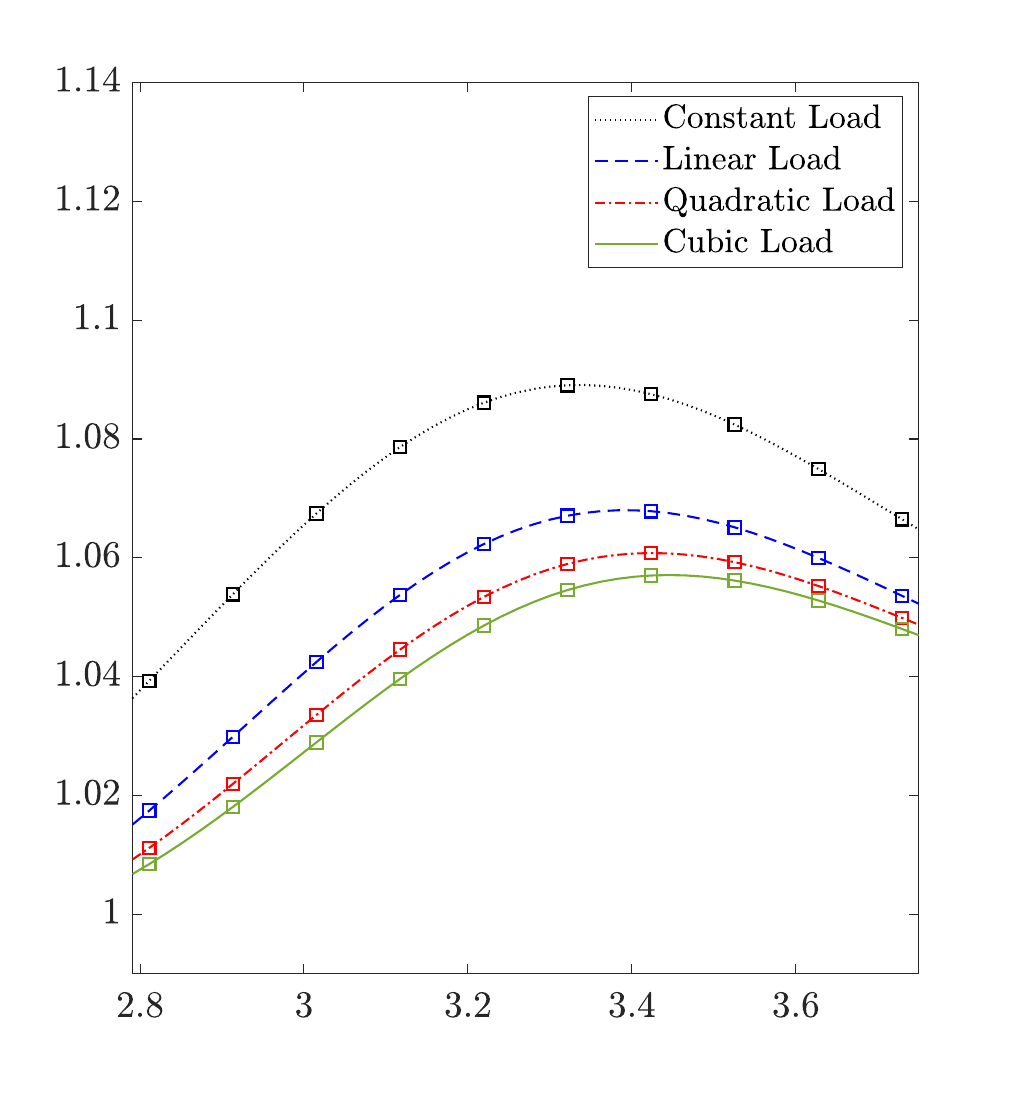}
  \put(-105,0) {$r$}
 \put(-235,115) {$\frac{w(r,1)}{w^*(r,1)}$}
 \put(-225,215) {{\bf (d)}}
 \put(-152,215) {{\it Larger crack - $\delta = 2/3$}}
 \caption{The ratio of the normalized displacement experienced at the top of a cylinder ($z=1$) with a crack, $w(r,1)$, against that experienced in the absence of a crack, $w^* (r,1)$. The crack is located at height $\delta$ and radial position between $2.75 < r < 3.75$. We show results for: {\bf (a)}, {\bf (c)} fixed linear loading on the inside surface $P_0 (z)$, and variable crack height $\delta$, {\bf (b)}, {\bf (d)} fixed $\delta=2/3$, and variable loading on the inside surface, for the case of a {\bf (a)}, {\bf (b)} smaller crack, {\bf (c)}, {\bf (d)} larger crack (see Table.~\ref{Table:CrackSize}). All remaining parameters are fixed as outlined in Table.~\ref{Table_Constants} for the symmetrical case. }
 \label{Fig:w}
\end{figure}

\begin{table}[b]
\centering
\begin{tabular}{c|c|c}
\hline \hline
Configuration & $c_0$ [m] & $c_1$ [m] \\
\hline
Smaller crack & $6\times 10^{-3}$ & $7 \times 10^{-3}$ \\
\hline
Larger crack & $5.9 \times 10^{-3}$ & $7.1 \times 10^{-3}$ \\
\hline \hline
\end{tabular}
\caption{The different sizes of crack, defined in terms of the inner and outer radius ($c_0$ and $c_1$ respectively), utilized in simulations in Sect.~\ref{Sect:Detect}. Note that both cracks are centrally located within the cylinder, while the `smaller crack' corresponds to the centrally located case considered in Sect.~\ref{Sect:CrackGrowth}.}
\label{Table:CrackSize}
\end{table}

Let us consider the effect of the crack on the displacement experienced at the top of the cylinder ($z=1$), again taking the case of a small steel cylinder. This point is chosen as it is the location where measurements are most easily taken. In order to account for different crack sizes we consider two different fracture geometries, which are outlined in Table.~\ref{Table:CrackSize}. All other parameters are taken in line with Table.~\ref{Table_Constants}.

The ratio of the normalized displacement experienced at this point, against that experienced in the absence of a crack, $w^*$, is provided in Fig.~\ref{Fig:w} for a variety of crack locations and torsion loadings (for solution $w^*$, see e.g. \cite{VAYSFELD2017526}, or this can also be obtained by inserting $\chi\equiv 0$ into the presented formulation). It can be seen that there is a clear quantitative effect of the crack on the displacement, with the crack leading to a small increase in displacement for fractures near to the upper surface, but a far larger increase in displacement when the crack is located in the bottom half of the cylinder. Although the crack location impacts the magnitude of the displacement, the outline of the crack is not directly visible within the observed displacement, even when it is near to the surface ($\delta=0.9$). 


\begin{figure}[t!]
 \centering
 \includegraphics[width=0.45\textwidth]{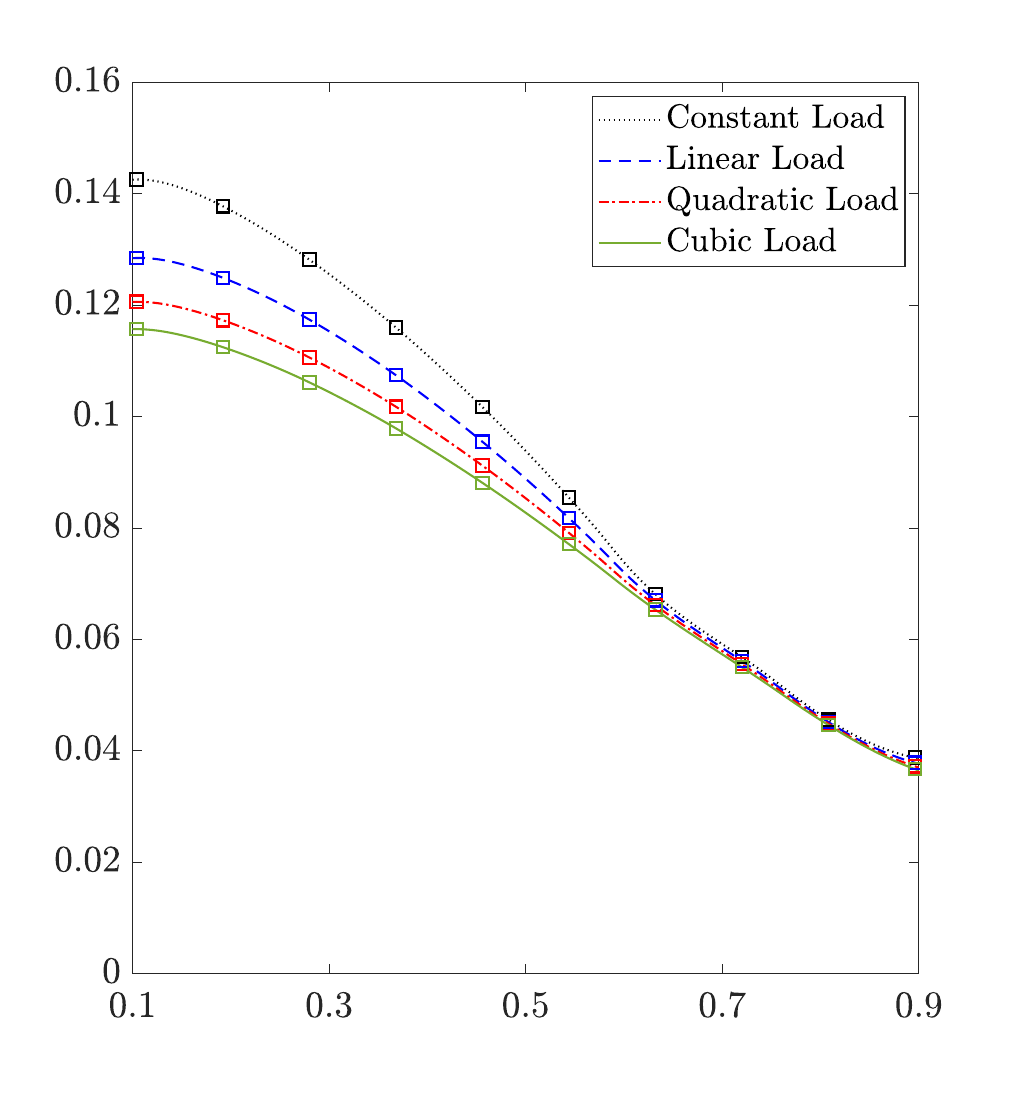}
  \put(-105,0) {$\delta$}
 \put(-235,115) {$f(\rho_1,\delta)$}
 \put(-225,215) {{\bf (a)}}
 \put(-135,215) {{\it Smaller crack}}
 \hspace{4mm} 
 \includegraphics[width=0.45\textwidth]{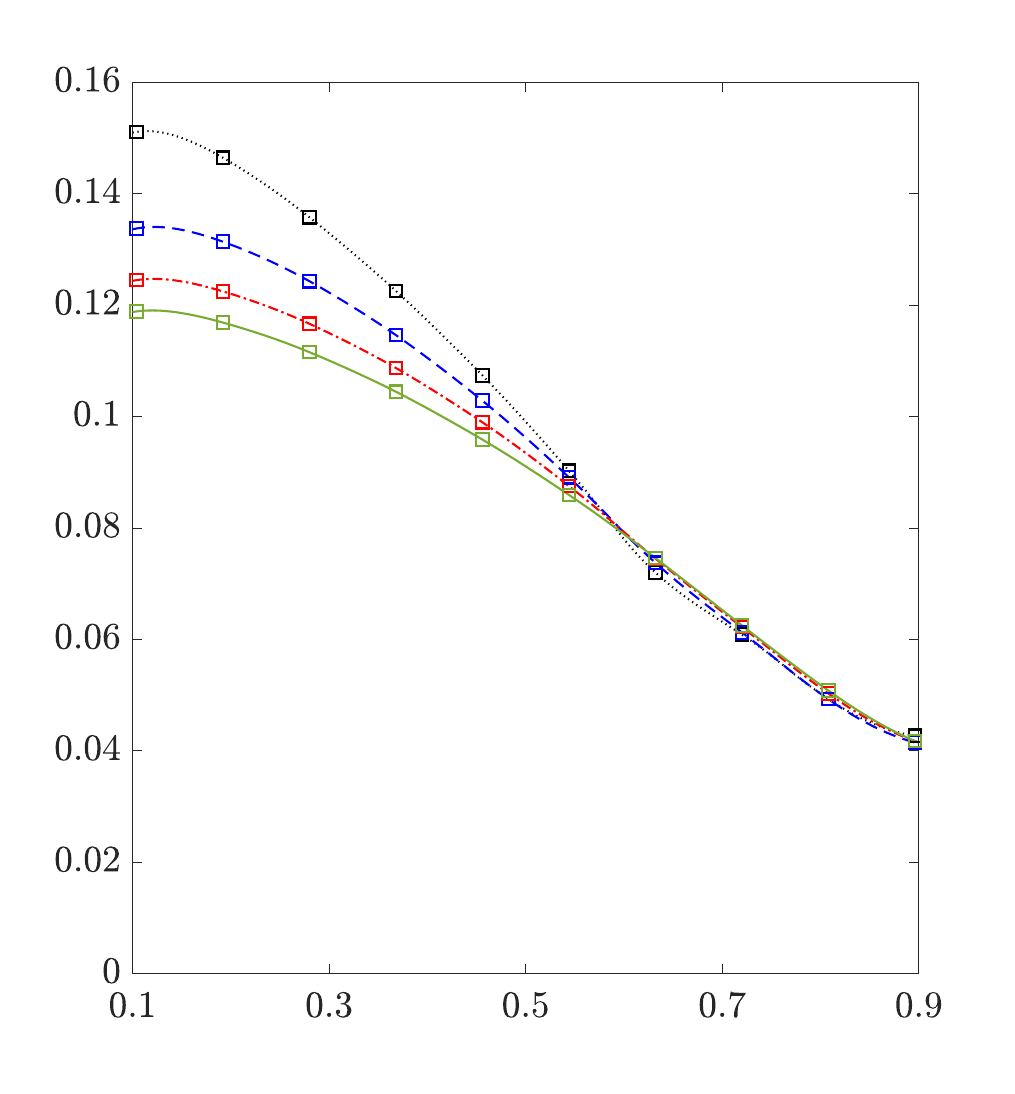}
  \put(-105,0) {$\delta$}
 \put(-235,115) {$f(\rho_1,\delta)$}
 \put(-225,215) {{\bf (b)}}
 \put(-130,215) {{\it Larger crack}}
 \caption{The relative influence of the crack on the normalized displacement  ${f}(\rho_1 , \delta)$ \eqref{f_rel_diff} experienced at the top of the cylinder ($z=1$) on the outer edge ($r=\rho_1$), as a function of the crack location (normalized height $\delta$) for various loading regimes. Figures show the case of a {\bf (a)}, {\bf (b)} smaller crack, {\bf (c)}, {\bf (d)} larger crack (see Table.~\ref{Table:CrackSize}). All remaining parameters are fixed as outlined in Table.~\ref{Table_Constants} for the symmetrical case. }
 \label{Fig:wBeta}
\end{figure}
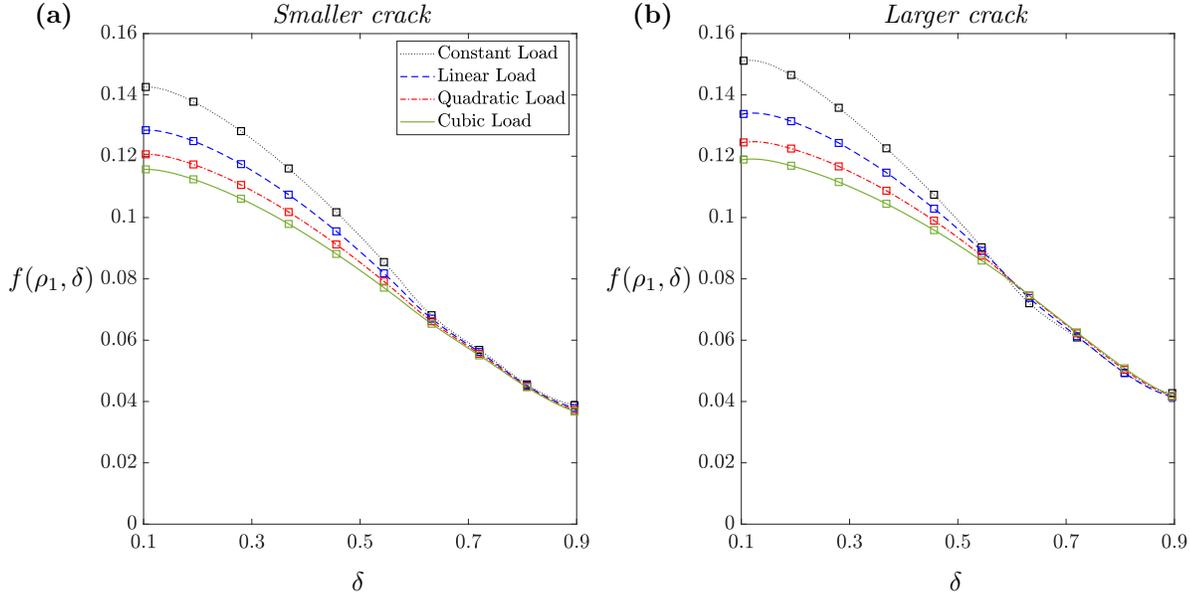

We therefore seek to use the quantitative impact of the crack on the displacement to determine its location. To this end, we evaluate the relative difference 
\begin{equation} \label{f_rel_diff}
f(r,\delta) = \frac{w(r,1;\delta)- w^* (r,1)}{w^*(r,1)} ,
\end{equation}
where $w$ is the normalized displacement for a given crack height $\delta$, while $w^*$ is the displacement in the absence of a crack. 

The value of $f(\rho_1, \delta)$, where $\rho_1$ is the outer edge of the cylinder, is provided in Fig.~\ref{Fig:wBeta} for various loadings. It can be seen that in the case of the larger crack there is an intersection of the curves for different loadings at approximately $\delta=0.6$, however there is no such intersection for the smaller crack. Furthermore, as the fracture approaches the surface ($\delta\to 1$), the curves for different loadings almost coincide, and remain closely packed together. These results, and their clear sensitivity to the crack size, mean that it would be difficult to use this measurement to determine the vertical position $\delta$ of a crack, and so we instead seek a clearer indicator.

In simulations it was found that, while the displacement experienced within the cylinder with a crack depends on $r$, the relative difference is almost constant with respect to $r$. 
Numerical investigations of $f(r,\delta)$ by the authors found that it varies by less than $1$\% over the problem domain for all fixed values of $\delta$ considered, irrespective of the loading applied. 

Consequently, we can average the relative difference $f(r,\delta)$ over $r$ and consider the difference solely in terms of $\delta$, yielding the difference function
\begin{equation} \label{f_delta}
\tilde{f}(\delta)= \frac{1}{\rho_1 -\rho_0} \int_{\rho_0}^{\rho_1} f(r,\delta) \, dr ,
\end{equation}
where $\rho_0$, $\rho_1$ are the normalized inner and outer edges of the cylinder respectively. This now provides an averaged measure of the cracks' influence on the normalized displacement. 

\begin{figure}[t!]
 \centering
  \includegraphics[width=0.45\textwidth]{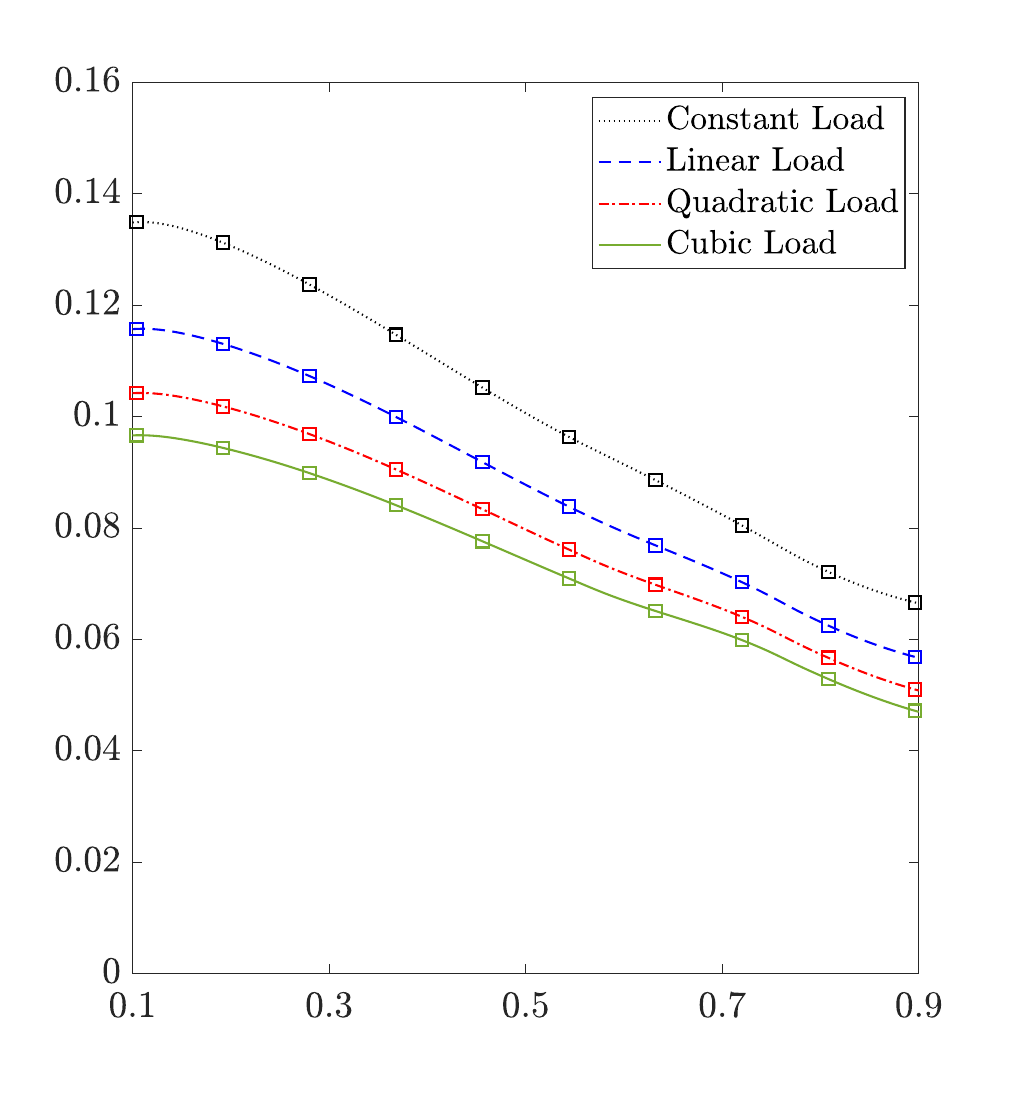}
  \put(-105,0) {$\delta$}
 \put(-225,115) {$\tilde{f}(\delta)$}
 \put(-225,215) {{\bf (a)}}
 \put(-135,215) {{\it Smaller crack}}
 \hspace{4mm} 
 \includegraphics[width=0.45\textwidth]{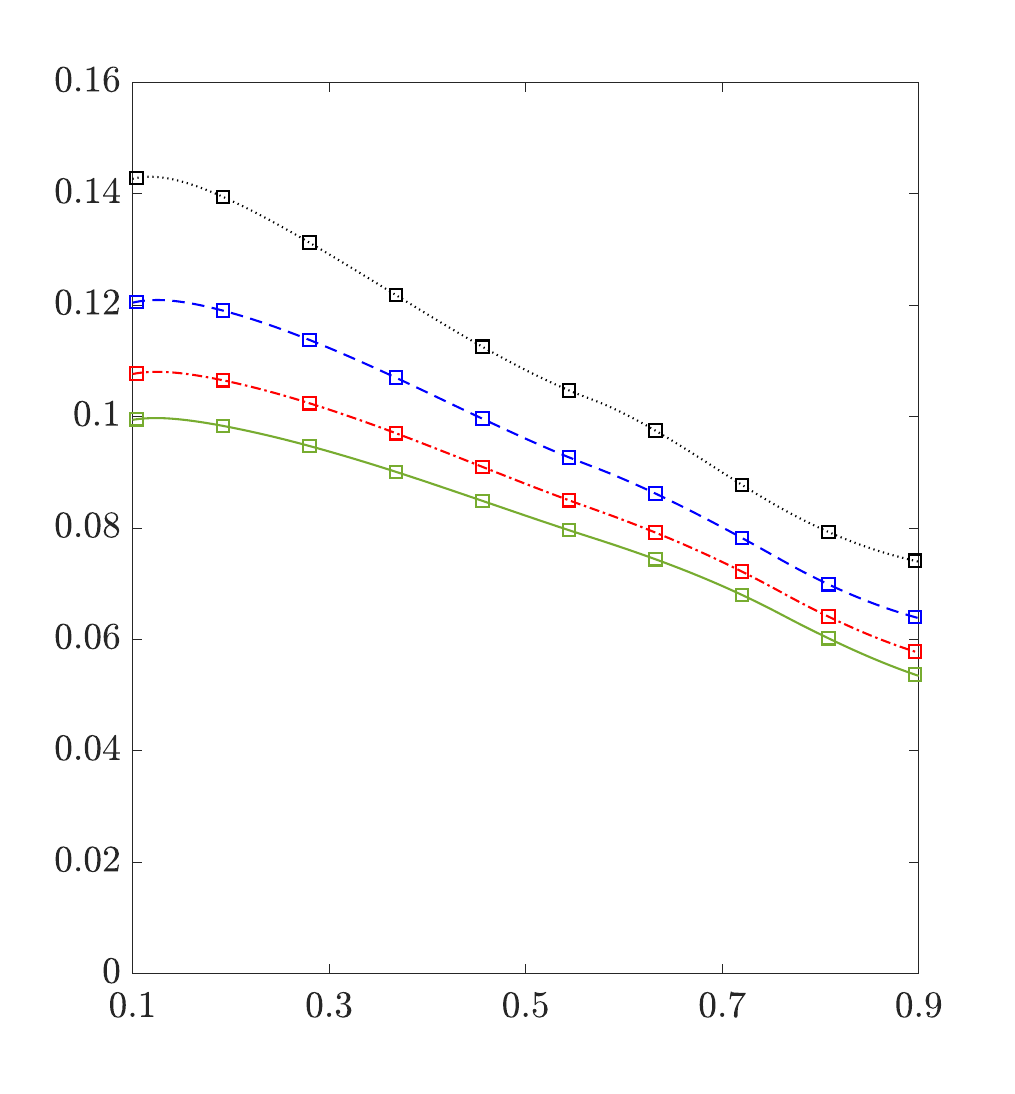}
  \put(-105,0) {$\delta$}
 \put(-225,115) {$\tilde{f}(\delta)$}
 \put(-225,215) {{\bf (b)}}
 \put(-130,215) {{\it Larger crack}}
 \caption{The relative influence of the cracks' presence on the normalized displacement, $\tilde{f}(\delta)$ \eqref{f_delta}, as a function of the location (normalized height $\delta$) of the crack within the cylinder. Figures show the case of a {\bf (a)}, {\bf (b)} smaller crack, {\bf (c)}, {\bf (d)} larger crack (see Table.~\ref{Table:CrackSize}). All remaining parameters are fixed as outlined in Table.~\ref{Table_Constants} for the symmetrical case.}
 \label{Fig:fDelta}
\end{figure}
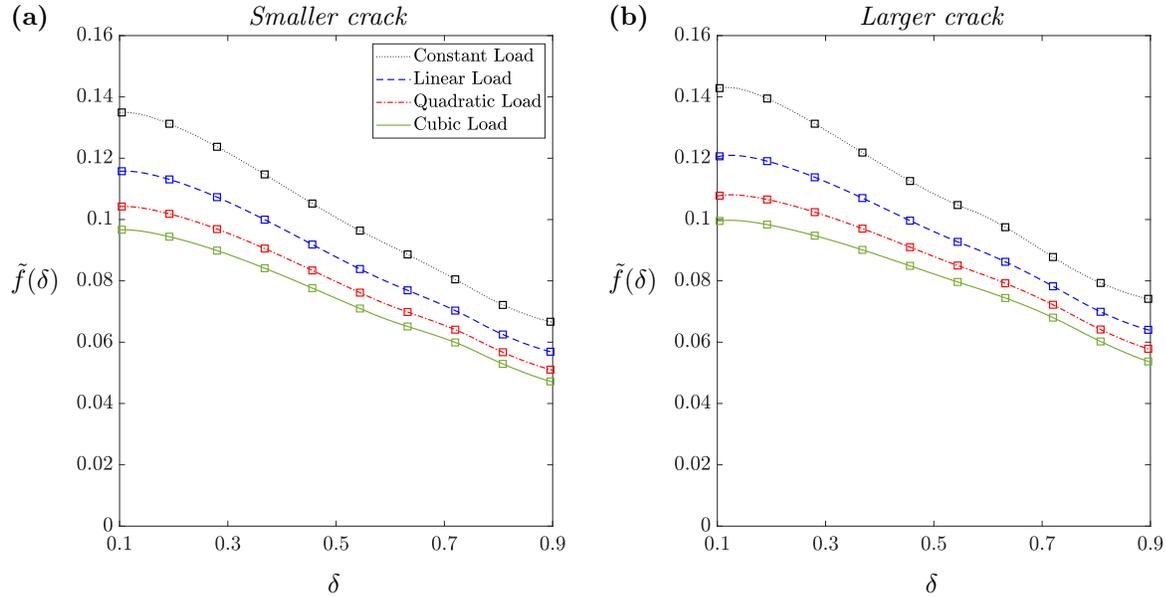

The results for $\tilde{f}(\delta)$ under a variety of applied loadings are provided in Fig.~\ref{Fig:fDelta}, for both the smaller and larger crack. It can be seen that the presence of the crack increases the displacement observed at the top of the cylinder. This effect decreases as the crack approaches the top of the cylinder, as there is less impacted material between the crack and the measured region. Over most of the region considered, the cracks' relative influence on the normalized displacement is almost linear, however we would not expect this to be the case as the crack approached the surface ($\delta\to 1$) in the considered formulation. The case of very small ($0<\delta \ll 1$) and large ($ 0 < 1-\delta \ll 1$) values of $\delta$ is left for future research.


One crucial aspect of $\tilde{f}(\delta)$ observed in Fig.~\ref{Fig:fDelta} is that, unlike the relative difference without averaging $f(r,\delta)$ \eqref{f_rel_diff} (see Fig.~\ref{Fig:wBeta}), there is no intersection between the curves representing different loading regimes. This result provides a test which can be performed in order to detect a crack within the cylinder and determine its approximate vertical location. The cylinder is placed under a variety of loading regimes (at least two, e.g. constant and linear), and the displacement at the top of the cylinder is measured in each case. The results are used to evaluate $\tilde{f}$ \eqref{f_delta}, as the average over $r$ of the relative difference between the measured displacement and the analytical solution in the absence of a crack. The obtained $\tilde{f}$ can then be compared with Fig.~\ref{Fig:fDelta} to determine whether a crack is present, and find the value of $\delta$ which best fits the results. 

To make this test effective, an approach to determining (or controlling for) the size of the crack needs to be determined. This is not only as it influences the value of $\tilde{f}$ observed in Fig.~\ref{Fig:fDelta}, but because the size of the crack, given in this formulation by normalized inner and outer radius $\alpha$ and $\beta$ respectively, was also shown to play a key role in the crack stress intensity factor (see Fig.~\ref{Fig:c0}). It is possible that this could be achieved by utilizing the analysis of $\tilde{f}$ (see Fig.~\ref{Fig:fDelta}), with the crack size being determined from the location of the intersection point for $f(r,\delta)$ (see Fig.~\ref{Fig:wBeta}), or directly from the displacement distribution (see Fig.~\ref{Fig:w}). Note as well that the presented results are for a steel cylinder, and accounting for material parameters will again add an additional layer of complexity.


Therefore, we can state that developing such a full test for the presence, location, and size of a crack within a cylinder requires a more detailed analysis. However, the presented results strongly indicate that one can be developed from the presented formulation, at least for a fixed cylinder geometry or material parameters.

\section{Concluding remarks}\label{Sect:Concl}

The problem of a rotating hollow cylinder containing a crack was considered. The form of the displacement and traction were obtained, and an iterative scheme developed to solve the resulting system.

Numerical investigations for the case of a small steel cylinder were undertaken. It was demonstrated that:
\begin{itemize}
 \item the presence of the crack has an influence on the displacement experienced by the cylinder, which is primarily dependent on the location of the crack;
 \item the stress intensity factor experienced on the crack surfaces is greater when the crack is located near to the edge of the cylinder (as opposed to being centrally located);
 \item the stress intensity factor was typically positively valued, and decreases with increasing crack size;
 \item rotation only played a negligible role for the small cylinder considered here, but may play a more prominent role for larger structures.
\end{itemize} 

The presented formulation can be used to predict whether crack extension will occur once the fracture location is known. The quasi-static formulation can be utilized to determine whether fracture initiation is an immediate concern, or it can be coupled with fatigue models (e.g. Paris' law) to estimate the extent of fatigue crack over time. These capabilities provide useful insight for risk management, for example when determining the number of cycles that can be `safely' performed on a cracked cylinder, or the number of cycles that can be performed between inspections.

Whether the presented formulation can be used to detect and locate cracks within a cylinder was also investigated.  The initial results indicate that the presence of a crack can be inferred from examining the displacement on the surface of the cylinder, and comparing the results with the case without a crack being present (see e.g. Fig.~\ref{Fig:w}). It was shown that, for a fixed crack size, examining the displacement under multiple loading configurations (at least two) and considering the average of the relative difference compared to the case without a crack can be used to determine crack height $\delta$ (see Fig.~\ref{Fig:fDelta}). 

This investigation was however conducted for a single cylinder geometry, and further analysis is required to produce a general procedure. Most notably, one which can determine both the crack location (height and radial position), and size, which are necessary to predict the likelihood of fracture extension over time (see Sect.~\ref{Sect:CrackGrowth}). One crucial advantage of such a test for crack location is that it can be conducted while the cylinder is undergoing rotation and loading (provided the prescribed loading can be induced), meaning that it could be employed while the cylinder is still `in use', rather than needing to stop and remove the component. Locating the crack position also allows for targeted repairs (where possible) to be planned and performed with minimal disruption.

%

When developing such a test for locating a fracture within the cylinder, it may also be useful to simplify the problem by replacing the crack by a soft interface (see e.g. \cite{OCHSNER2007703,Mishuris2004}) across the entire cross-section. Comparison of the results with the presented formulation could be used to ensure accuracy of the analysis. This approximated problem may be used to develop a simplified test for locating the crack, but can also be utilized to examine additional important effects, such as the case of an imperfect interface (impact of force conservation)/ the edge effect \cite{MISHURIS2005409}. Finally, the presence of a damage zone within the neighbourhood of the existing fracture can also be incorporated \cite{Mishuris2001a,Mishuris2001b}, to better model the impact of the crack on the cylinder behaviour. 



\section*{Acknowledgements}

The research is supported by European project funded by Horizon 2020 Framework Programme for Research and  Innovation (2014-2020) (H2020-MSCA-RISE-2020)       Grant  Agreement   number 101008140 EffectFact “Effective Factorisation techniques for matrix-functions: Developing theory, numerical methods and impactful applications”. 

\bibliography{Hidden_Crack_Bib}
\bibliographystyle{abbrv}

\end{document}